\documentclass[final,11pt,sort&compress,times]{elsarticle}

\usepackage{amssymb}

\journal{Journal of Computational Physics}

\usepackage{amsmath}
\usepackage{stmaryrd}
\usepackage{hyperref}

\renewcommand{\d}{{\rm d}}
\newcommand{\jp}{{j+\frac{1}{2}}}
\newcommand{\jm}{{j-\frac{1}{2}}}
\newcommand{\jpc}{{j+1/2}}
\newcommand{\jmc}{{j-1/2}}
\newcommand{\ten}[1]{\mathsf{#1}}
\newcommand{\jump}[1]{{\llbracket {#1} \rrbracket}}

\begin{document}

\begin{frontmatter}



\title{Well balancing of the SWE schemes for moving-water steady flows}

\author[EndIF_adr]{Valerio Caleffi\corref{cor1}}
\ead{valerio.caleffi@unife.it}
\cortext[cor1]{Corresponding author}
\author[EndIF_adr]{Alessandro Valiani}
\ead{alessandro.valiani@unife.it}
\address[EndIF_adr]{Department of Engineering, University of Ferrara,
Via G. Saragat, 1, 44122 Ferrara, Italy}

\begin{abstract}
In this work, the exact reproduction of a moving-water steady flow via the numerical solution of the one-dimensional shallow water equations is studied.

A new scheme based on a modified version of the HLLEM approximate Riemann solver [Dumbser M. and Balsara D.S., J. Comput. Phys. 304 (2016) 275--319] that exactly preserves the total head and the discharge in the simulation of smooth steady flows and that correctly dissipates mechanical energy in the presence of hydraulic jumps is presented. This model is compared with a selected set of schemes from the literature, including models that exactly preserve quiescent flows and models that exactly preserve moving-water steady flows.

The comparison highlights the strengths and weaknesses of the different approaches. In particular, the results show that the increase in accuracy in the steady state reproduction is counterbalanced by a reduced robustness and numerical efficiency of the models. Some solutions to reduce these drawbacks, at the cost of increased algorithm complexity, are presented.

\end{abstract}


\end{frontmatter}

\section{Introduction}\label{sec:intro}
In recent years, the attention of several researchers involved in
studying the numerical discretization of the classical nonlinear
shallow water equations (SWE) has been focused on the exact preservation of specific asymptotic flow states.

In particular, two asymptotic states over a non-flat bottom are
considered: the first one consists of the motionless steady state
that is characterized by a constant free-surface elevation and a zero
specific discharge, and the second one consists of the moving steady flow, which is characterized by the constancy of the total head and of the specific discharge \cite{Chow}. The attention devoted to the exact reproduction of these steady cases is because some common flows can be interpreted as small perturbations of such asymptotic states \cite{BeVa-94,YSN2011,Cheng2016}. 

In the case of quiescent flows, the preservation of such a state is
related to the correct balancing between the flux gradients and the
bottom-slope source term. Therefore, the numerical schemes that
exactly reproduce this asymptotic case are denoted as \emph{well
  balanced} (WB). Following the work of Berm\'udez and
V\'azquez-Cend\'on \cite{BeVa-94}, a numerical model that is able to
exactly preserve an initial quiescent flow is also denoted as
\emph{C-property} satisfying. An updated review on this topic can be
found in \cite{XS14}. At present, for several scheme families,
different techniques can be found in the literature to achieve the
well balancing. 

A popular approach for the well-balancing problem is developed in the
context of the balance laws in non-conservative form
\cite{DLM,Gosse01,Pares2004,PCL,Castro2007}. In general, studying the
balance laws in the non-conservative form is complicated by the
difficulty in rigorously defining the correct weak solution if a
discontinuity is present. To address this problem, the DLM theory
proposed by Dal Maso et al. \cite{DLM} represents a very powerful
tool. In the DLM theory, a family of paths that link the conserved
variables across the discontinuity is selected to define the weak
solution. The concept of using a family of paths to address the
solution discontinuity is used by Par\'es in \cite{PCL} to construct
the \emph{path-conservative} (or path-consistent) schemes. The
application of the path-conservative schemes to the SWE is performed
by Castro et al. in \cite{Castro2007}. The first key element of this
approach is enriching the classical SWE by a trivial equation obtained by
equating to zero the time derivative of the bottom elevation. In this
augmented system of equations, the source term related to the bottom
slope is written as a non-conservative product, and the entire system
is therefore in the non-conservative form
\cite{Gosse01,Castro2007}. This augmented formulation allows the
conservative part of the SWE and the source term related to the bottom
elevation to be solved in a unified manner. Castro et
al. \cite{Castro2007} show that the use of a linear path leads to
well-balanced schemes without further complications. Note that the
eigenvalues and the eigenvectors associated with the augmented SWE are
very simple, and this allows very efficient numerical models to be developed.

The same formalism of the path-conservative schemes is adopted by
Dumbser and Toro in \cite{DOT} to extend the Riemann solver of Osher
and Solomon \cite{Osher1982} to certain classes of hyperbolic systems
in non-conservative form, leading to the DOT Riemann solver. This
path-conservative Osher-type scheme has several advantages with
respect to other path-conservative schemes. It does not require an
entropy fix, it is very simple to implement, and it can be considered
a complete Riemann solver \cite{Toro09}. The application of the DOT
Riemann solver to the augmented SWE, together with a linear path,
leads to a well-balanced scheme \cite{DOT,Caleffi2015a}. This scheme,
denoted in the following as the PC model, is taken into account in
this work for comparison with other methods.

Revisiting classical solutions for the Riemann problem (RP), Dumbser
and Balsara \cite{Dumbser2016} introduce a new simple and very general
formulation of the approximate solver HLLEM. The HLLEM Riemann solver,
originally proposed in \cite{Einfeldt1988} and \cite{Einfeldt1991},
has interesting properties, such as robustness, the positivity
preserving property and entropy enforcement. In the new formulation
presented in \cite{Dumbser2016}, these properties are preserved while
the resolution of the linearly degenerated waves is greatly improved
at the cost of the computation of the only eigenvalues and
eigenvectors related to the linearly degenerated fields. Moreover, in
\cite{Dumbser2016}, a formulation of the HLLEM solvers for a
non-conservative system of equations is proposed. Again, the formalism
of the path-conservative schemes is adopted to extend the approach to
the non-conservative balance laws. The use of a segment (linear) path
in the definition of the scheme applied to the augmented SWE leads to
a scheme that satisfies the C-property. This last form of the HLLEM solver is considered in this work for the comparison.

In the PC and HLLEM models, the source term is taken into account by
increasing the number of equations involved in the mathematical
model. Following a dual approach, in \cite{Murillo2010}, the source
term is accounted for by increasing the number of elementary waves in
the Riemann solver used to construct a Roe-type scheme. In the
classical formulation of the Roe Riemann solver \cite{ROE}, the
approximate RP solution is obtained by exactly solving a suitable linearization of the homogeneous part of the balance laws. Consequently, the approximate solution of the system with $m$ equations is constituted by $m$ waves that separate $m+1$ constant states. The source term is discretized apart.
In the augmented Roe solver (ARoe) \cite{Murillo2010}, the source term
is discretized together with the homogeneous part of the balance law
and, by again using a suitable linearization, the approximated
solution composed of $m+1$ waves and $m+2$ constant states is
achieved. The application of these concepts to the SWE allows
obtaining an RP approximate solution to be used in a Godunov-type
scheme. The corresponding model, denoted as the ARoe model herein, is
well balanced.

To the best of our knowledge, the first scheme developed to exactly
preserve a moving-water steady state was proposed by Noelle et
al. \cite{noelle2007}. In this work, a high-order accurate finite
volume scheme, which is exactly well balanced for the general steady
flow, is presented. In the case of an asymptotic steady state, this
scheme is able to preserve the total head where the flow is smooth
while the correct amount of energy dissipation is obtained in the
hydraulic jumps. From the perspective of energy conservation and
dissipation, this scheme can be denoted as \emph{energy balanced} (EB) \cite{Murillo2013}.

Following the pioneering work of Noelle et al. \cite{noelle2007},
several studies have been devoted to analyzing the energy balancing of the schemes. The reader is referred to  \cite{LeFloch2011,FjMiTa2011,noelle2007,Murillo2013,xing2014,Caleffi2015a,Cheng2016,Hiltebrand2016} for a non-exhaustive list of references. These works address both theoretical and practical aspects of the problem.

The importance of exactly preserving a moving-water steady state is
highlighted in \cite{YSN2011,Cheng2016}. In particular, the advantages
of the energy-balanced schemes over the well-balanced schemes are shown
in \cite{YSN2011}. In this work, by analyzing the numerical reproduction of small perturbations of steady flows, the superiority of the EB methods over the WB methods is proven.

The main theoretical analysis of the exact preservation of meaningful
asymptotic states is conducted in the general context of the balance
laws. For example, Tadmor \cite{Tadmor2016} studies the problem in
terms of an entropy function using the concepts of
\emph{entropy-conservative} and \emph{entropy-stable} schemes
introduced in \cite{tadmor1987}. These analyses can easily be applied
to the SWE taking into account that the total energy plays the role of
an entropy function in the context of the shallow water equations with
a bottom source term. In fact, in the case of a steady flow governed
by the SWE, it is important to distinguish the cases characterized by
the presence or absence of a hydraulic jump. If the solution is
continuous (i.e., a hydraulic jump is not present), then the
asymptotic case is characterized by the uniformity of the specific
discharge and the total head. In this case, the scheme is called
\emph{energy preserving} \cite{Murillo2013}. When a hydraulic jump is
present and the correct amount of mechanical energy is dissipated, the
appropriate definition of the scheme is \emph{energy consistent}
\cite{Murillo2013}. In \cite{FjMiTa2011}, the stability analysis of an
SWE numerical model in terms of entropy (total energy) is specifically
addressed for the first time, and the corresponding model is entropy
stable. Working in the same framework, different authors propose
entropy-stable models. For example, in \cite{Hiltebrand2016}, a
shock-capturing space-time discontinuous Galerkin method that is
energy stable, C-property satisfying, and able to cope with unstructured grids is presented.

Murillo and Garc\'ia-Navarro \cite{Murillo2013}, in the context of the augmented Roe solvers \cite{Murillo2010}, propose a strategy to extend the well-balanced property for the quiescent state of the ARoe scheme to the moving-water steady state. This approach is based on a revisited integration of the bed source term for the discontinuous bottom. We have considered this extension, and we denote the corresponding model as the ARoeEB scheme.

In the framework of the path-conservative schemes, a general
definition of a well-balanced scheme for a generic asymptotic state is
given in \cite{Pares2004,Castro2013}. In these works, a numerical
method is defined as \emph{well balanced} for a given integral curve related to a linearly degenerated vector field if, given any steady solution that belongs to that integral curve and an initial discrete state that belongs to the same integral curve,  the initial state is exactly preserved \cite{Pares2004,Castro2013}. 
For the particular case of the SWE \eqref{eq:PCSWE}, the definition of
the well-balanced scheme proposed in \cite{Pares2004,Castro2013} and
the definition of the energy-balanced scheme used in
\cite{Murillo2013} are coincident because it is easy to show that the
generalized Riemann invariants associated with the integral curve of
the linearly degenerated field are the specific discharge and the
total head \cite{Caleffi2015a}. Working in the context described in
\cite{Pares2004,Castro2013}, in \cite{Caleffi2015a}, a new
path-consistent and energy-balanced scheme is obtained by combining a
suitable curvilinear path with a DOT Riemann solver \cite{DOT}. The
corresponding numerical scheme, denoted as the PCEB model herein, is considered for the comparison with the other schemes.

The good results obtained using the curvilinear path in the DOT
framework \cite{Caleffi2015a} have motivated the authors to
investigate the use of the same approach in different schemes. Taking
into account that the HLLEM model is developed using the same
formalism of the path-conservative schemes, the application of a
non-linear path follows in a natural manner. Therefore, in this work,
we present an original formulation of the HLLEM solver, denoted as the
HLLEMEB method herein, that allows the energy balancing of the scheme
for moving-water steady flows to be achieved. 

Another classical approach developed to achieve the C-property
satisfaction that can be extended to the energy-balanced scheme is the
hydrostatic reconstruction \cite{audusse04}. In \cite{audusse04}, a
flux correction to be applied to a Godunov-type finite volume scheme
is introduced to achieve the well balancing. Following the same idea of a suitable flux correction, Caleffi and Valiani \cite{Caleffi2009a} suggest a strategy to achieve the exact solution of a steady flow. This method is based on a suitable reconstruction of the conservative variables at the cell interfaces, coupled with a correction of the numerical flux based on the local conservation of total energy. For this reason, we can denote this approach as a \emph{hydrodynamic reconstruction}. A first-order implementation of this scheme is considered in this work and is denoted as the HDEB model.

Finally, it is interesting to note that several approaches available
in the literature assume a discontinuous bottom topography. In this condition, the augmented SWE \cite{LeFloch2011,Pares2004,Castro2013} is not a strictly hyperbolic system of equations, and this allows the occurrence of the resonance phenomenon  \cite{LeFloch2011}. This situation is achieved when the characteristic speeds become coincident and the solution of a Riemann problem may be not unique or can be constituted by a larger number of elementary waves with respect to the non-resonant solution. Understanding the behavior of the numerical models in the resonant regime is an important aspect for validating each numerical scheme. 

In this work, together with the presentation of the new
energy-balanced scheme HLLEMEB, we present a comparison between a set
of schemes from the literature, including models that exactly preserve
quiescent flows and models that exactly preserve moving-water
steady flows. The seven selected schemes are the PC, PCEB, HLLEM,
HLLEMEB, ARoe, ARoeEB and the HDEB schemes. The comparison highlights
the strengths and weaknesses of the different approaches. The comparison shows that the better reproduction of the steady flows, or small perturbation of such steady flows, is counterbalanced by a loss of numerical efficiency and robustness of the models. Some solutions to reduce these drawbacks, at the cost of increased algorithm complexity, are presented.

The remainder of this paper is organized as follows. In section
\ref{sec:math_model}, the SWE mathematical model is presented in both
the conservative and non-conservative forms. The eigenvalues and
eigenvectors, together with the relevant physical quantities
associated with the SWE, are introduced. In section
\ref{sec:numerical}, the description of each considered model is
presented. Each description is well detailed to improve the
reproducibility of the results. A large space is devoted to the new
HLLEMEB model, which is presented here for the first time. In section
\ref{sec:comparison}, detailed comparisons between the models are
described. To show the behavior of the different models in the
reproduction of the resonant regime, a specific Riemann problem is
also used as a test case. Finally, some conclusions are drawn.

\section{The mathematical model}\label{sec:math_model}
We consider the classical nonlinear shallow water equations \cite{Chow}. To simplify the comparison between models, only the source term related to the bottom topography is taken into account, while the source term related to the friction is neglected. Under this assumption, the SWEs become:
\begin{equation}\label{eq:SWE}
\partial_t u + \partial_x f = s; \quad \text{with:} \quad u=\begin{bmatrix}h\\ q \end{bmatrix};
\quad f=\begin{bmatrix}q\\ \frac{gh^2}{2}+\frac{q^2}{h}\end{bmatrix};
\quad s=\begin{bmatrix}0\\ -g\,h\,z_x\end{bmatrix};
\end{equation}
where $u(x,t)$ is the vector of the conservative variables, $f(u)$ is
the flux vector, $s(u,x)$ is the source term vector, $h(x,t)$ is the
water depth, $q(x,t)$ is the specific discharge, $z(x)$ is the bottom
elevation, $g$ is the gravity acceleration, and $x$ and $t$ are the space and time, respectively.

To apply the theoretical framework of the path-conservative schemes \cite{PCL}, the SWE can be written as a quasi-linear PDE system introducing the trivial equation $z_t=0$ \cite{Castro2013}:
\begin{equation}\label{eq:PCSWE}
\partial_t w + A(w)\,\partial_x w = 0; \quad \text{with:} \quad w=\begin{bmatrix}h\\ q \\ z \end{bmatrix}; \quad
A(w)=\begin{bmatrix}0&1&0\\c^2-v^2&2v&c^2\\0&0&0\end{bmatrix};
\end{equation}
where $w(x,t)$ is the augmented vector of the conservative variables, $v=q/h$ is the depth-averaged velocity, and $c=\sqrt{g\,h}$ is the wave celerity. The matrix $A$ has the following eigenvalues:
\begin{equation}\label{eq:eigenvalues}
\lambda_1=v-c; \quad \lambda_2=0; \quad \lambda_3=v+c;
\end{equation}
and, indicating the Froude number with $\ten{Fr}=|v|/c$, the following right eigenvectors:
\begin{equation}\label{eq:eigenvectors}
R_1 = \begin{bmatrix}1\\ \lambda_1 \\ 0 \end{bmatrix}; \quad
R_2 = \begin{bmatrix}1\\ 0 \\ \ten{Fr}^2-1 \end{bmatrix}; \quad
R_3 = \begin{bmatrix}1\\ \lambda_3 \\ 0 \end{bmatrix}.
\end{equation}

The three eigenvalues \eqref{eq:eigenvalues} are collected in the diagonal matrix $\Lambda=\text{diag}(\lambda_1,\lambda_2,\lambda_3)$, and the three eigenvectors \eqref{eq:eigenvectors} are collected in a matrix $R = [R_1,R_2,R_3]$ (i.e., the $j$-th column of the matrix $R$ is the vector $R_j$).

In the following, we extensively use the specific energy, $E$:
\begin{equation}\label{eq:defE}
E = h+\frac{q^2}{2\,g\,h^2};
\end{equation}
and the total head, $H$ \cite{Chow}:
\begin{equation}\label{eq:defH}
H = z+E.
\end{equation}

\section{The numerical models}\label{sec:numerical}
To improve the reproducibility of this study, an accurate description of the models is provided in this section. 
In particular, a large space is devoted to explaining the HLLEMEB
model, which is presented here for the first time. To focus the
attention on the comparison between the different approaches, we only consider the simplest first-order implementations without the complexities related to the high-order versions. However, the extension to high-order accuracy can be performed using standard techniques. All the schemes are explicit in time, and a standard CFL condition is used \cite{Toro09}.
\subsection{The PC model - A DOT path-conservative model}\label{sec:pcl}
In this section, the standard DOT path-conservative model, which is
well balanced only for the quiescent flow, is described. 

Integrating Eq.~\eqref{eq:PCSWE} over the cell $I_j = \left[x_\jmc,x_\jpc \right]$, the fundamental equation for a first-order path-conservative scheme is \cite{PCL}:
\begin{equation} \label{eq:update}
w_j^{n+1} = w_j^{n} - \frac{\Delta t}{\Delta x} \left[ \mathcal{D}^{-}_{\jp} + \mathcal{D}^{+}_{\jm} \right];
\end{equation} 
where $w_j^{n+1}$ and $w_j^{n}$ are the cell-averaged values of the vector $w$ at time levels $t^{n+1}$ and $t^{n}$, respectively. The fluctuations $\mathcal{D}^{\pm}_{\jpc}$, generally depending on the discontinuous values $w_{\jpc}^{-}$ and $w_{\jpc}^{+}$ at the cell interfaces $x_\jpc$, are computed using the Dumbser-Osher-Toro (DOT) Riemann solver \cite{DOT}:
\begin{equation} \label{eq:Dpm}
\mathcal{D}^{\pm}_{\jp} = \frac{1}{2}\int_0^1 \left[
A(\Psi(w_{\jp}^{-},w_{\jp}^{+},s)) \pm 
\left|A(\Psi(w_{\jp}^{-},w_{\jp}^{+},s)) \right|\right]\frac{\partial \Psi}{\partial s}\ \d s
\end{equation} 
where the absolute-value matrix-operator is defined by $|A| =
R\,|\Lambda|\, R^{-1}$, with
$|\Lambda|=\text{diag}(|\lambda_1|,|\lambda_2|,|\lambda_3|)$. $\Psi(w_{\jpc}^{-},w_{\jpc}^{+},s)$
is the integration path, which is typically given as a parametrized function of $s \in [0,1]$. 
In the case of a first-order accurate model, we have the identities $w_{\jpc}^{-}=w_{j}^n$ and $w_{\jpc}^{+}=w_{j+1}^n$.

To complete the description of this model, we must introduce an explicit expression for the path $\Psi(w_{\jpc}^{-},w_{\jpc}^{+},s)$ to be inserted in Eq.~\eqref{eq:Dpm}.

To understand the necessity of the integration path, recall that the
presence of non-conservative products in Eq.~\eqref{eq:PCSWE} makes
the definition of a correct weak solution difficult in the case of
discontinuities of the bottom profile or of the solution itself
\cite{DLM,PCL}. A well-established approach for overcoming this
difficulty is proposed by Dal Maso et al. \cite{DLM} in the so-called
DLM theory. In this theory, by introducing a suitable family of paths
that link the conservative variables across a discontinuity, the definition of the weak solution becomes possible (see the original work by Dal Maso et al. \cite{DLM} for details of the theory).
%
%
%
%
%
%
Here, we recall only that in the DLM theory must the admissible path satisfy a generalized Rankine-Hugoniot condition for shocks in the following form \cite{DLM,ALOUGES2004}:
\begin{equation}\label{eq:RH_generalized}
\int_0^1 A(\Psi(w^-,w^+,s)) \frac{\partial \Psi}{\partial s} \d s = \zeta \left(w^{+} - w^{-}\right),
\end{equation}
where $\zeta$ is the celerity of the shock and $w^{-}$ and $w^{+}$ are
the values of the variables before and after the interface, respectively. 

Unfortunately, the DLM theory does not provide information about the choice of the path for general balance laws.
Notwithstanding, several proposals for defining the physically correct
path exist (see \cite{ALOUGES2004} and the references therein). Among
these proposals, the vanishing viscosity approach introduced in
\cite{LeFloch1990} is well established. In this framework, the physically correct path that should be inserted in Eq~\eqref{eq:RH_generalized} for a general system of the form of \eqref{eq:PCSWE} can be selected studying the traveling wave (i.e., the viscous profile) associated with the modified system of equations:
\begin{equation}\label{eq:PCSWE_Viscous}
\partial_t w^{\varepsilon} + A(w)\,\partial_x w^{\varepsilon} = \varepsilon \partial_x\left(B(w^{\varepsilon})\partial_x w^{\varepsilon} \right);
\end{equation}
with a Heaviside initial condition, where $w^{\varepsilon}$ is the solution of the system \eqref{eq:PCSWE_Viscous}, $B$ is a physically based viscosity matrix, and $\varepsilon$ is the viscosity parameter. The admissible solution of \eqref{eq:PCSWE} is the vanishing viscosity limit of the solution of \eqref{eq:PCSWE_Viscous} \cite{LeFloch1990,LeFloch2014,Berthon2015}. More precisely, the solution $w$ of the system \eqref{eq:PCSWE} is considered admissible for a given matrix $B$ if $w= \lim_{\varepsilon \to 0} w^{\varepsilon}$ (almost everywhere). 
Consequently, some mathematical manipulations allow showing that the correct path connecting $w^-$ and $w^+$ must be the solution $w^{\varepsilon}$ after a suitable re-parameterization \cite{LeFloch1990,LeFloch2014}.
This approach requires the definition of a viscosity matrix $B$ on a
physical basis, and this is not always easy. Moreover, the form of
this matrix has to be rather simple to obtain a system of the form of
Eq.~\eqref{eq:PCSWE_Viscous} that can be analytically solved for and a
Heaviside initial condition. Some attempts to obtain this matrix in the family of the shallow water models are performed in \cite{Berthon2015}, but a definitive answer on the nature of $B$ for the SWE with a bottom discontinuity is not provided. 

In this work, following the idea introduced in several works \cite[e.g.:][]{PCL,Pares2004,DOT,Dumbser2016}, rather simple paths are used in our numerical model, and the correctness of the choice is verified \emph{a posteri}. In this manner, we preserve some degree of freedom in the selection of the path to obtain a simple implementation of the model and specific behaviors of the numerical solutions.

In particular, working on the SWE, the use of a very simple linear path is sufficient to obtain reasonable results if only the motionless steady state has to be preserved \cite{Pares2004}. In this case, indicating with $\jump{\star}$ the jump of the generic variable $\star$ across the cell interface (i.e., $\jump{\star}=\star^{+}-\star^{-}$), the path $\Psi(w^-,w^+,s)$
is defined as:
\begin{equation}\label{eq:linpathexp}
\Psi(w^-,w^+,s) = w^- + s(w^+ - w^-) = w^- + s\jump{w};
\end{equation}
or, explicitly:
\begin{equation}
\Psi(s) = \begin{bmatrix}
\bar{h}(s)   \\ 
\bar{q}(s)   \\
\bar{z}(s)   \end{bmatrix}=
\begin{bmatrix}
h^-_\jp + s \jump{h_\jp}  \\ 
q^-_\jp + s \jump{q_\jp}  \\
z^-_\jp + s \jump{z_\jp}   \end{bmatrix}.
\nonumber
\end{equation}
Eq.~\eqref{eq:linpathexp} is inserted in Eq.~\eqref{eq:Dpm}, and the integration is performed numerically. A Gaussian quadrature with 3 points yields satisfactory results; see \cite{DOT} for details. The use of Eq.~\eqref{eq:linpathexp} in Eq.~\eqref{eq:Dpm} leads to a well-balanced model for a quiescent flow \cite{Pares2004}.
%
%
\subsection{The PCEB model - A DOT path-conservative model with a nonlinear path}\label{sec:pcn}
The use of the segment path does not allow an initial moving-water
steady state to be exactly preserved \cite{Caleffi2015a}. For this reason, a different path, inspired by  M\"{u}ller and Toro \cite{Mueller13} and Par{\'{e}}s and Castro \cite{Castro2013}, is introduced in \cite{Caleffi2015a}.

To justify the nature of the new path, we must take into account the role of the integral curve related to the linearly degenerated vector field in the definition of a well-balanced scheme for a general steady flow.
As mentioned in the introduction, a numerical method is well balanced
for a given integral curve related to a linearly degenerated vector
field if any initial steady solution that belongs to that integral
curve is exactly preserved at the discrete level
\cite{Pares2004,Castro2013}. This general definition, when applied to
the context of the SWE, can be formulated assuming that a model is
energy balanced if an initial steady condition that is characterized by a constant specific discharge and a constant total head is preserved at the discrete level \cite{Castro2013}. 

With some algebra, it is simple to show that the use of a path in \eqref{eq:Dpm}
that, in steady conditions, is a parametrization of an arc of the integral curve related to the linearly degenerated field allows obtaining an energy-balanced model in the sense of \cite{Pares2004,Castro2013}. The details can be found in \cite{Caleffi2015a}.

In the particular case of the SWE, the quantities that are invariant along the integral curve (i.e., the generalized Riemann invariants) are the specific discharge and the total head. Therefore, it is natural to parametrize the integral curve (and therefore the selected path) in terms of specific discharge and total head.

A path $\Psi(w^-,w^+,s)$ that in the case of steady state corresponds to an arc of the integral curve is:
\begin{equation}\label{eq:pathnl}
\Psi(s) = \begin{bmatrix}
\bar{h}(s)   \\ 
\bar{q}(s)   \\
\bar{z}(s)   \end{bmatrix}=
\begin{bmatrix}
\bar{E}(s)^{-1}  \\ 
q^-_\jp + s (q^+_\jp - q^-_\jp)   \\
z^-_\jp + s (z^+_\jp - z^-_\jp)   \end{bmatrix};
\end{equation}
with:
\begin{equation}\label{eq:Espec}
\bar{E}(s) = \bar{h}(s)+\frac{[\bar{q}(s)]^2}{2\,g\,[\bar{h}(s)]^2} = \bar{H}(s) - \bar{z}(s);
\end{equation}
and:
\begin{equation}\label{eq:pathHlin}
\bar{H}(s) = \bar{H}^-_{\jp} + s (\bar{H}^+_{\jp} - \bar{H}^-_{\jp}).
\end{equation}
The computation of $\bar{E}^{-1}$, i.e., finding the values of $\bar{h}$ that satisfy \eqref{eq:Espec}, is not trivial. In fact, Eq.~\eqref{eq:Espec} for given values of $\bar{H}$, $\bar{q}$ and $\bar{z}$ is a cubic equation in $\bar{h}$ that is generally solved numerically. In this work, to obtain a computationally efficient scheme, we have used the analytical exact solution given in \cite{Valiani2008}. 

Note that the trivial introduction of the path defined by
Eq.~\eqref{eq:pathnl} into Eq.~\eqref{eq:Dpm} leads to a
computationally expensive model. However, the availability of the
analytical expressions for the eigenvalues, the eigenvectors and
$\bar{E}^{-1}$ allows analytical simplifications to be performed that
lead to a very efficient computation of the fluctuations. In
\ref{apx:explicit}, we show the fully explicit expressions to be used
for computing the fluctuations \eqref{eq:Dpm}.

The path \eqref{eq:pathnl} works correctly only in the cases where
both $w^-$ and $w^+$ are subcritical or supercritical. In the case of a transcritical flow, better performance is obtained using a path divided into three parts, as described in \ref{sec:path3s}.

\subsection{The HLLEM model}\label{sec:HLLEM}
We treat the approximate Riemann solver HLLEM for systems in non-conservative form, recently proposed by Dumbser and  Balsara \cite{Dumbser2016}. This approach can be considered as an extension of the classical HLL solver \cite{HLL}. In this work, we present the formulation of the scheme specific for the SWE, and the details about the formulation for general balance laws can be found in \cite{Dumbser2016}.

To introduce the HLLEM Riemann solver, the SWEs are written in the following form:
\begin{equation}\label{eq:SWEHLLEM}
\partial_t w + \partial_x \hat{f}+ B(w)\,\partial_x w = 0;
\end{equation}
with:
\begin{equation}
\quad w=\begin{bmatrix}h\\ q \\ z \end{bmatrix}; \quad
\hat{f}(w)=\begin{bmatrix}q\\ \frac{gh^2}{2}+\frac{q^2}{h}\\0\end{bmatrix};\quad 
B(w)=\begin{bmatrix}0&0&0\\0&0&c^2\\0&0&0\end{bmatrix};
\end{equation}
where the genuinely conservative part of the system is represented by the non-linear flux $\hat{f}(w)$ and the non-conservative part is represented by $B(w)\,\partial_x w$. Clearly, we have the relationship $\partial_w \hat{f}+B=A$.
Note that only one element of the matrix $B$ is different from zero; therefore, every computation that involves $B(w)\,\partial_x w$ or its derived quantities are in general computationally cheap.
To simplify the computation, Eq.~\eqref{eq:SWEHLLEM} is rewritten introducing the similarity variable $\xi=x/t$, following the approach also used in \cite{Balsara2010,Balsara2014,Balsara2015}:
\begin{equation}\label{eq:SWEHLLEMSIM}
w- \partial_\xi (\xi w) + \partial_\xi \hat{f}+ B(w)\,\partial_\xi w = 0;
\end{equation}

Because the HLL Riemann solver \cite{HLL} is used as a starting point for the development of the HLLEM solver, we start our description from the reinterpretation of the HLL solution.
In the original HLL Riemann solver, the approximate solution is strongly simplified. In particular, the solution is constituted by one intermediate state $w_*$ separated from the unperturbed states $w^-$ and $w^+$ by the fastest outward-moving waves:
\begin{equation}\label{eq:hll}
w(\xi)=\left\{\begin{array}{l}
w^-,\ \text{if}\ \xi\leq s^-,\\
w_*,\ \text{if}\ s^-<\xi< s^+, \\
w^+,\ \text{if}\ \xi\geq s^+ \end{array} \right.
\end{equation}
where $s^-$ and $s^+$ are the celerity estimates of the two outward waves. See \S~\ref{sec:wave_estimates} for a brief discussion about the computation of the outward wave celerities.

Taking into account the solution \eqref{eq:hll}, the discontinuities at $s^-$ and $s^+$, the presence of the non-conservative product in \eqref{eq:SWEHLLEMSIM} and the DLM theory \cite{DLM}, the integration of \eqref{eq:SWEHLLEMSIM} between $s^-$ and $s^+$ can be written as:
\begin{multline}\label{eq:intmp}
w_* (s^+-s^-) - (w^+\,s^+ - w^-\,s^-) + (\hat{f}^+ - \hat{f}^-) + \\
+ \int_0^1 B(\Psi(w^-,w_*,s))\frac{\partial \Psi}{\partial s}\d s +
\int_0^1 B(\Psi(w_*,w^+,s))\frac{\partial \Psi}{\partial s}\d s =0.
\end{multline}
where $\hat{f}^\pm = \hat{f}(w^\pm)$. Introducing the segment path \eqref{eq:linpathexp} into \eqref{eq:intmp}, we obtain:
\begin{multline}\label{eq:intmp1}
w_* (s^+-s^-) - (w^+\,s^+ - w^-\,s^-) + (\hat{f}^+ - \hat{f}^-) + \\
+ \tilde{B}(w^-,w_*)(w^* - w^-) + \tilde{B}(w_*,w^+)(w^+ - w^*) =0.
\end{multline}
with:
\begin{equation}\label{eq:tildeB}
\tilde{B}(w_a,w_b) = \int_0^1 B(\Psi(w_a,w_b,s))\d s = \begin{bmatrix}0&0&0\\0&0&\frac{1}{2}g(h_a + h_b)\\0&0&0\end{bmatrix}.
\end{equation}
It is important to note the very simple expression of the matrix $\tilde{B}$. Only one element is different from zero, and its computation is straightforward. To avoid matrix-vector products, we can introduce the following vector:
\begin{equation}
\hat{b}(w_a,w_b) = \tilde{B}(w_a,w_b) (w_b - w_a) = \begin{bmatrix}0\\ \frac{1}{2}g(h_a + h_b) (z_b-z_a) \\0\end{bmatrix}
\end{equation}

Now, from \eqref{eq:intmp1}, we can write:
\begin{equation}\label{eq:wstar}
w_* = \frac{
(w^+\,s^+ - w^-\,s^-) - (\hat{f}^+ - \hat{f}^-) - \left( \hat{b}(w^-,w_*) + \hat{b}(w_*,w^+) \right)}{(s^+-s^-) }.
\end{equation}
Eq.~\eqref{eq:wstar} represents an implicit expression of $w_*$ that can be solved numerically starting from the guess value, $w^0_*$, given explicitly by:
\begin{equation}\label{eq:wstar0}
w^0_* = \frac{
(w^+\,s^+ - w^-\,s^-) - (\hat{f}^+ - \hat{f}^-) - \hat{b}(w^-,w^+)}{(s^+-s^-)}.
\end{equation}
The computed $w_*$ is a key element of the HLL solver.

The fluctuations associated with the HLL scheme are obtained by integrating \eqref{eq:SWEHLLEMSIM} in $[s^-,0]$ and $[0,s^+]$:
\begin{align}
\int_{s^-}^{0} \left(w- \partial_\xi (\xi w) \right) \d \xi + \int_{s^-}^{0}\left(\partial_\xi \hat{f}+ B(w)\,\partial_\xi w \right) \d \xi= 0;\\
\int_{0}^{s^+} \left(w- \partial_\xi (\xi w) \right) \d \xi+ \int_{0}^{s^+}\left(\partial_\xi \hat{f}+ B(w)\,\partial_\xi w \right) \d \xi= 0.
\end{align}
The second terms on the left-hand sides of the above equations correspond to the numerical fluctuations; therefore, we can write:
\begin{align}\label{fluctHLLcomp1D}
\mathcal{D}^{-}_{\text{HLL}}(w^-,w^+) &= \int_{s^-}^{0} \left(w- \partial_\xi (\xi w) \right) \d \xi;\\\label{fluctHLLcomp2D}
\mathcal{D}^{+}_{\text{HLL}}(w^-,w^+) &= \int_{0}^{s^+} \left(w- \partial_\xi (\xi w) \right) \d \xi.
\end{align}

The right-hand sides are known quantities if we assume that $w_*$ is computed using \eqref{eq:wstar}. In fact, the substitution of \eqref{eq:wstar} in \eqref{fluctHLL1}-\eqref{fluctHLL2} leads to:
\begin{align}\label{fluctHLL1}
\mathcal{D}^{-}_{\text{HLL}}(w^-,w^+) &= - \frac{s^-}{s^+ - s^-} \hat{\mathcal{P}}(w^-,w^+,w_*) + \frac{s^-s^+}{s^+ - s^-} (w^+ - w^-);\\
\mathcal{D}^{+}_{\text{HLL}}(w^-,w^+) &= + \frac{s^+}{s^+ - s^-} \hat{\mathcal{P}}(w^-,w^+,w_*) - \frac{s^-s^+}{s^+ - s^-} (w^+ - w^-);\label{fluctHLL2}
\end{align}
with:
\begin{equation}\label{fluctHLL3}
\hat{\mathcal{P}}(w^-,w^+,w_*) = \hat{f}^+ - \hat{f}^- +\hat{b}(w^-,w_*) +\hat{b}(w_*,w^+).
\end{equation}

It is well known that the resolution of the contact discontinuity
related to the presence of linearly degenerated fields by the HLL
scheme is poor \cite{Toro09} due to an excessive numerical
diffusion. To improve the behavior of the scheme, an
\emph{anti-diffusive} term is introduced. As suggested in
\cite{Einfeldt1988,Einfeldt1991}, to extend the HLL Riemann solver to
take the presence of the linearly degenerated intermediate waves into account, the constant intermediate state $w_*$ is replaced with a piecewise linear state:
\begin{equation}\label{eq:hllem}
w(\xi)=\left\{\begin{array}{ll}
w^-   &\text{if}\ \xi\leq s^-,\\
w_* + \varphi R_*(\bar{w})2\delta_*(\bar{w})L_*(\bar{w})\frac{w^+ w^-}{s^+ s^-}
\left(\xi - \frac{1}{2}(s^-+s^+)\right)   &\text{if}\ s^-<\xi< s^+, \\
w^+   &\text{if}\ \xi\geq s^+ \end{array} \right.
\end{equation}
where $\bar{w}=\left(w^-+w^+\right)/2$; $R_*$ and $L_*$ are the
matrices of the right and left eigenvectors associated with the
linearly degenerated fields, respectively; $\varphi$ is a
regularization parameter that allows smoothly switching between the
HLL solver and the HLLEM solver; and $\delta_*$ is given by:
\begin{equation}
\delta_* = I - \frac{\Lambda_*^{-}}{s^-} - \frac{\Lambda_*^{+}}{s^+};
\end{equation}
where $I$ is the identity matrix and $\Lambda_*$ is the diagonal
matrix containing the eigenvalues associated with the linearly
degenerated fields and $\Lambda_*^{\pm} = \frac{1}{2} (\Lambda_* \pm
|\Lambda_*|)$. See \cite{Dumbser2016} for further details.

Following the procedure for computing the HLL fluctuations, the HLLEM
fluctuations can be computed by substituting \eqref{eq:hllem} into:
\begin{align}\label{fluctHLLcomp1}
\mathcal{D}^{-}_{\text{HLLEM}}(w^-,w^+) &= \int_{s^-}^{0} \left(w- \partial_\xi (\xi w) \right) \d \xi\\\label{fluctHLLcomp2}
\mathcal{D}^{+}_{\text{HLLEM}}(w^-,w^+) &= \int_{0}^{s^+} \left(w- \partial_\xi (\xi w) \right) \d \xi;
\end{align}

obtaining, after some algebraic manipulations:
\begin{align}\label{fluctHLLEMp1}
\mathcal{D}^{-}_{\text{HLLEM}}(w^-,w^+) &= \mathcal{D}^{-}_{\text{HLL}}(w^-,w^+) + \nonumber \\
&- \varphi \frac{s^-s^+}{s^+ - s^-} R_*(\bar{w})\delta_*(\bar{w})L_*(\bar{w})(w^+ - w^-);\\
\label{fluctHLLEMp2}
\mathcal{D}^{+}_{\text{HLLEM}}(w^-,w^+) &= \mathcal{D}^{+}_{\text{HLL}}(w^-,w^+) + \nonumber \\
&+\varphi \frac{s^-s^+}{s^+ - s^-} R_*(\bar{w})\delta_*(\bar{w})L_*(\bar{w})(w^+ - w^-).
\end{align}

In the specific case of the SWE \eqref{eq:SWEHLLEM}, also considering \eqref{eq:eigenvectors}, it is:
\begin{equation}
R_*=R_2=\begin{bmatrix}1\\ 0 \\ \bar{\ten{Fr}}^2-1 \end{bmatrix}; \quad
L_*=\begin{bmatrix}0\quad 0\quad \frac{1}{\bar{\ten{Fr}}^2-1}\end{bmatrix};\quad 
\delta_* = 1;
\end{equation}
where $\bar{\ten{Fr}}$ is the Froude number computed for $\bar{w}$. Defining $\mathcal{T}$ as:
\begin{equation}\label{eq:antidiff}
\mathcal{T}(w^-,w^+) = R_*(\bar{w})\delta_*(\bar{w})L_*(\bar{w})(w^+ - w^-) = \begin{bmatrix}\frac{1}{\bar{\ten{Fr}}^2-1} (z^+ - z^-)\\ 0 \\ (z^+ - z^-) \end{bmatrix}
\end{equation}
and assuming $\varphi=1$, the HLLEM fluctuations become:
\begin{align}\label{fluctHLLEM1}
\mathcal{D}^{-}_{\text{HLLEM}}(w^-,w^+) &= \mathcal{D}^{-}_{\text{HLL}}(w^-,w^+) 
- \frac{s^-s^+}{s^+ - s^-} \mathcal{T}(w^-,w^+);\\
\label{fluctHLLEM2}
\mathcal{D}^{+}_{\text{HLLEM}}(w^-,w^+) &= \mathcal{D}^{+}_{\text{HLL}}(w^-,w^+) 
+ \frac{s^-s^+}{s^+ - s^-} \mathcal{T}(w^-,w^+).
\end{align}

The computation of the HLLEM fluctuations can be summarized in the
following steps: an estimate of $w_*$ is obtained iteratively from
Eq.~\eqref{eq:wstar} using $w_*^0$ given by \eqref{eq:wstar0} as a guess value; 
the HLL fluctuations $\mathcal{D}^{\pm}_{\text{HLL}}$ are computed using \eqref{fluctHLL1}-\eqref{fluctHLL3};
the \emph{anti-diffusive} term $\mathcal{T}(w^-,w^+)$ \eqref{eq:antidiff} is evaluated; finally, the HLLEM fluctuations $\mathcal{D}^{\pm}_{\text{HLLEM}}$ are computed using \eqref{fluctHLLEM1}-\eqref{fluctHLLEM2}.

The third components of the HLLEM fluctuations, which are related to
the bottom behavior, are always zero. To demonstrate this fact, it is
sufficient to note that the third components of $\hat{f}$ and
$\hat{b}$ are null by construction, and therefore, the third component
of $\hat{\mathcal{P}}$ in Eq.~\eqref{fluctHLL3} is also always
zero. Finally, the third components of the second terms of
\eqref{fluctHLL1}-\eqref{fluctHLL2} exactly cancel out the third
component of the last terms in \eqref{fluctHLLEM1}-\eqref{fluctHLLEM2}.
In other words, regarding the bottom elevation, the anti-diffusive term exactly balances the numerical diffusion introduced by the HLL scheme. 

As shown in \cite{Dumbser2016}, this scheme satisfies the C-property.

\subsubsection{Wave speed estimates}\label{sec:wave_estimates}
The HLLEM solver is quite sensitive to the estimates of $s^-$ and $s^+$. In \cite{Dumbser2016}, for general systems of conservative laws, the following approximations are suggested:
\begin{equation}
s^- = \min\left(0,\Lambda(w^-),\Lambda(\bar{w}) \right);
\quad \text{and} \quad
s^+ = \max\left(0,\Lambda(w^+),\Lambda(\bar{w}) \right);
\end{equation}
where $\bar{w}=\left(w^-+w^+\right)/2$. In this work, this simple approach does not provide optimal results, and therefore, we have used an indirect approach based on the \emph{two-rarefactions approximation} \cite{toro01}.

An approximate intermediate state is computed assuming that the elementary waves that separate the intermediate state from the unperturbed states are rarefactions. Under this assumption, the intermediate state is characterized by:
\begin{equation}
v_* = \frac{1}{2}(v^-+v^+) + c^- - c^+;
\quad \text{and} \quad
c_* = \frac{1}{2}(c^- + c^+) + \frac{1}{4}(v^- - v^+);
\end{equation}
where $v_*$ and $c_*$ are the velocity and the celerity in the intermediate state, respectively. The wave celerities become:
\begin{equation}
s^- = \min\left(0,v^- - c^-,v_* - c_* \right);
\quad \text{and} \quad
s^+ = \max\left(0,v^+ + c^+,v_* + c_* \right).
\end{equation}
\subsection{The HLLEMEB model - An energy-balanced HLLEM model}
To improve the behavior of the HLLEM scheme in the case of a steady
solution, we modify the computation of the fluctuations. An original
energy-balanced model, denoted here as HLLEMEB, is obtained in this way.

First, we denote with $\hat{J}$ the Jacobian matrix of the flux $\hat{f}$, i.e.,  $\hat{J}= \partial_w \hat{f}$, and we recall that:
\begin{equation}
\int_0^1 \hat{J}(\Psi(w_a,w_b,s))\frac{\partial \Psi}{\partial s}\d s = \hat{f}_b - \hat{f}_a;
\end{equation}
independently from the chosen path. Clearly, we also have $\hat{J} + B = A$. Consequently, the integration of \eqref{eq:SWEHLLEMSIM} between $s^-$ and $s^+$ can be written as:
\begin{equation}\label{eq:intmpWB}
w_* (s^+-s^-) - (w^+\,s^+ - w^-\,s^-) + \mathcal{P}(w^-,w^+,w_*) = 0.
\end{equation}
where:
\begin{equation}\label{eq:mathcalP}
\mathcal{P}(w^-,w^+,w_*) = \int_0^1 A(\Psi(w^-,w_*,s))\frac{\partial \Psi}{\partial s}\d s +
\int_0^1 A(\Psi(w_*,w^+,s))\frac{\partial \Psi}{\partial s}\d s.
\end{equation}

Eq.~\eqref{eq:intmpWB} represents a different formulation of Eq.~\eqref{eq:intmp}.
Straightforward manipulations also lead to:
\begin{equation}\label{eq:wstarWB}
w_* = \frac{(w^+\,s^+ - w^-\,s^-) - \mathcal{P}(w^-,w^+,w_*)}{s^+-s^-}.
\end{equation}
Eq.~\eqref{eq:wstarWB} represents an implicit expression for $w_*$.

To obtain an energy-balanced scheme, we require that if $w^-$ and
$w^+$ belong to the same integral curve related to the linearly
degenerate field (i.e., are characterized by the same specific
discharge and total head), then  $w_*$ also belongs to the same integral curve. In general, this property is not exactly respected by \eqref{eq:wstarWB}; thus, we introduce a suitable path and some small corrections.

First, we use the same non-linear path of the scheme described in \S~\ref{sec:pcn} defined by the relationship \eqref{eq:pathnl}-\eqref{eq:pathHlin}. This choice becomes particularly interesting when $w^-$, $w^+$ and $w_*$ belong to the same integral curve related to the linearly degenerate field.  In fact, taking into account the relationships in \ref{apx:explicit}, and in particular Eq.~\eqref{eq:Adpds}, it is easy to show that Eq.~\eqref{eq:mathcalP} becomes:
\begin{equation}\label{eq:mathcalPExp}
\mathcal{P}(w^-,w^+,w_*) = 
\begin{bmatrix}
(q^+-q^-)\\
I_{v1} (q_*-q^-) + I_{c1} (H_*-H^-) + I_{v2} (q^+ - q_*) + I_{c2} (H^+ - H_*)\\
0
\end{bmatrix}
\end{equation}
with:
\begin{align}
I_{v1}&=\int_{w^-}^{w_*} v \ \d w;& I_{v2}&=\int_{w_*}^{w^+} v \ \d w;\\
I_{c1}&=\int_{w^-}^{w_*} c^2 \ \d w;& I_{c2}&=\int_{w_*}^{w^+} c^2 \ \d w;
\end{align}
and $\mathcal{P}$ clearly becomes zero in the case of a steady flow with $q^-=q^+=q_*$ and $H^-=H^+=H_*$. 

The use of the curvilinear path is useful but not sufficient; in fact,
if we assume treating a steady state characterized by $\bar{q}$ and $\bar{H}$, i.e., $q^\pm=\bar{q}$ and $z^\pm + h^\pm + (q^\pm)^2/[2g(h^\pm)^2] = \bar{H}$,  and to have $\mathcal{P} = 0$, we obtain from \eqref{eq:wstarWB} the relationships (written by components):
\begin{align}
h_* &= \frac{(h^+\,s^+ - h^-\,s^-)}{s^+-s^-};\\
q_* &= \frac{(q^+\,s^+ - q^-\,s^-)}{s^+-s^-};\\
z_* &= \frac{(z^+\,s^+ - z^-\,s^-)}{s^+-s^-};
\end{align}
that show that we have $q_* = \bar{q}$ but not $h_*=\bar{H}^{-1}$. Therefore, for steady states, $w^\pm$ from Eq.~\eqref{eq:wstarWB} does not lead to a consistent steady state $w_*$.

To correct this behavior, we introduce $H_{eb}$ as:
\begin{equation}\label{eq:hwb}
H_{eb} = \frac{(H^{+}\,s^+ - H^{-}\,s^-)}{s^+-s^-}.
\end{equation}
and we compute $\tilde{h}$ inverting the total head $H_{eb}$ assuming as given $q_*$ and $z_*$ (i.e., that are the second and third components of $w_*$).
Then, $h_{eb}$ can computed as:
\begin{equation}\label{eq:hstarWB}
h_{eb} = \tilde{h} - \frac{\mathcal{P}^{(1)}(w^-,w^+,w_*)}{s^+-s^-}.
\end{equation}
where $\mathcal{P}^{(1)}$ is the first component of the vector
$\mathcal{P}$. We also introduce $h_{nb}$ given by:
\begin{equation}
h_{nb} = \frac{(h^+\,s^+ - h^-\,s^-) - \mathcal{P}^{(1)}(w^-,w^+,w_*)}{s^+-s^-},
\end{equation}
which corresponds to the unmodified first component of \eqref{eq:wstarWB}. 
The depth $h_{eb}$ satisfies the energy-preserving requirement for a steady flow, while $h_{nb}$ is the correct value of the depth for a general unsteady flow. Note that $\mathcal{P}^{(1)}(w^-,w^+,w_*) = q^+ - q^-$, and therefore, it does not depend on $w_*$. Thus, this computation is not numerically expensive.

For a smooth transition between $h_{nb}$ and $h_{eb}$, we introduce a weight $\phi$:
\begin{equation}\label{eq:phi}
\phi = \tanh\left(\alpha \left|H^{+} - H^{-}\right| \right);
\end{equation}
with $\alpha=10$ (the results are not sensitive to the value of $\alpha$), and we compute $h_*$ as:

\begin{equation} \label{eq:hweight}
h_* = \phi\,h_{nb} + (1-\phi)\,h_{eb}.
\end{equation}

The substitution of \eqref{eq:hweight} in the first component of \eqref{eq:wstarWB}, taking into account Eqs.~\eqref{eq:hwb}-\eqref{eq:phi}, provides a system of equations to be solved numerically to obtain $h_*$, $q_*$ and $z_*$. The guess value is again computed using Eq.~\eqref{eq:wstar0}.
With these corrections, in the case of a steady flow, the numerical solution of \eqref{eq:wstarWB} converges to a proper solution.

Additionally, the \emph{anti-diffusive} term \eqref{eq:antidiff} has to be modified. To this end, we introduce the following coefficient:
\begin{equation}
C = \frac{h^{-}+h^{+}}{2\sqrt{h^{-}h^{+}}};
\end{equation}
and the new term:
\begin{equation}\label{eq:antidiffWB}
\tilde{\mathcal{T}}(w^-,w^+) = \begin{bmatrix}\frac{(z^+ - z^-)}{C(\ten{Fr}^{-}\ten{Fr}^{+})-1} \\ 0 \\ (z^+ - z^-) \end{bmatrix}.
\end{equation}

Now, the other elements of the HLLEMEB model are analogous to the corresponding elements of the HLLEM model. Following exactly the same procedure presented in the previous section \S~\ref{sec:HLLEM}, the HLL-like fluctuations are computed as:
\begin{align}\label{fluctHLLWB1}
\mathcal{D}^{-}_{\text{HLLEB}}(w^-,w^+) &= - \frac{s^-}{s^+ - s^-} \mathcal{P}(w^-,w^+,w_*) + \frac{s^-s^+}{s^+ - s^-} (w^+ - w^-);\\
\mathcal{D}^{+}_{\text{HLLEB}}(w^-,w^+) &= + \frac{s^+}{s^+ - s^-} \mathcal{P}(w^-,w^+,w_*) - \frac{s^-s^+}{s^+ - s^-} (w^+ - w^-);\label{fluctHLLWB2}
\end{align}
with $\mathcal{P}(w^-,w^+,w_*)$ given by \eqref{eq:mathcalPExp}. Clearly, Eqs.~\eqref{fluctHLLWB1}-\eqref{fluctHLLWB2} are the counterparts in the HLLEMEB model of Eqs.~\eqref{fluctHLL1}-\eqref{fluctHLL2} in the HLLEM model.

Finally, the energy-preserving fluctuations are computed as:
\begin{align}\label{fluctHLLEMEB1}
\mathcal{D}^{-}_{\text{HLLEMEB}}(w^-,w^+) &= \mathcal{D}^{-}_{\text{HLLEB}}(w^-,w^+) 
- \frac{s^-s^+}{s^+ - s^-} \tilde{\mathcal{T}}(w^-,w^+);\\
\label{fluctHLLEMEB2}
\mathcal{D}^{+}_{\text{HLLEMEB}}(w^-,w^+) &= \mathcal{D}^{+}_{\text{HLLEB}}(w^-,w^+) 
+ \frac{s^-s^+}{s^+ - s^-} \tilde{\mathcal{T}}(w^-,w^+);
\end{align}
which are the HLLEMEB counterparts of Eqs.~\eqref{fluctHLLEM1}-\eqref{fluctHLLEM2} in HLLEM.

The computation of the HLLEMEB fluctuations can be summarized in the
following steps: an estimate of $w_*$ is obtained iteratively from
Eq.~\eqref{eq:wstarWB} using $w_*^0$ of \eqref{eq:wstar0} as a guess
value; the HLL fluctuations $\mathcal{D}^{\pm}_{\text{HLLEB}}$ are
computed using \eqref{fluctHLLWB1}-\eqref{fluctHLLWB2}; the
\emph{anti-diffusive} term $\tilde{\mathcal{T}}(w^-,w^+)$ is evaluated
by \eqref{eq:antidiffWB}; and finally, the HLLEMEB fluctuations $\mathcal{D}^{\pm}_{\text{HLLEMEB}}$ are computed using \eqref{fluctHLLEMEB1}-\eqref{fluctHLLEMEB2}.

\subsection{The ARoe model - an augmented Roe model} \label{sec:ARoe}
The augmented Roe model (ARoe) is based on the approximate solution of the Riemann problem (RP) defined by the shallow water equations \eqref{eq:SWE} and a piecewise initial condition \cite{Murillo2010}:
%
%
\begin{align}
&\partial_t u + \partial_x f = s; \label{eq:RP1}\\
&u(x,0)=\left\{\begin{array}{ll}
u_{j}^{n}, &\text{if}\ x \leq x_\jp;\\
u_{j+1}^{n}, &\text{if}\ x > x_\jp; \end{array} \right.\label{eq:RP2}
\end{align}
where $u_j^{n+1}$ and $u_j^{n}$ are the cell-averaged values of the vector $u$ at the time levels $t^{n+1}$ and $t^{n}$, respectively.


Because the classical Roe approximate solution is used to define the
ARoe method, we recall here the key elements of the Roe solver
\cite{Toro09}. The classical Roe solution is obtained by exactly solving the linearized RP:
\begin{align}
&\partial_t u + \tilde{J}_\jp(u^{n}_{j},u^{n}_{j+1}) \partial_x u = s;\\
&u(x,0)=\left\{\begin{array}{ll}
u_{j}^{n}, &\text{if}\ x \leq x_\jp;\\
u_{j+1}^{n}, &\text{if}\ x > x_\jp; \end{array} \right.
\end{align}
where $\tilde{J}_\jpc(u^{n}_{j},u^{n}_{j+1})$ is a suitable constant Jacobian matrix for the homogeneous system. In particular, $\tilde{J}_\jpc(u^{n}_{j},u^{n}_{j+1})$ must be diagonalizable with real eigenvalues (i.e., the approximate linear system must be hyperbolic) and $\tilde{J}_\jpc(u^{n}_{j},u^{n}_{j+1}) \to \tilde{J}_\jpc(u^{n}_{j})$ smoothly as $u^{n}_{j+1} \to u^{n}_{j}$ (the linearized system must be consistent with the original SWE) \cite{Toro09}. We use a well-established technique to construct a valid linearized Jacobian matrix by introducing the Roe averages \cite{Toro09}:
\begin{equation}
c_\text{Roe}=\sqrt{g\frac{h^{n}_{j}+h^{n}_{j+1}}{2}}; \qquad v_\text{Roe}=\frac{v^{n}_{j}\sqrt{h^{n}_{j}} + v^{n}_{j+1}\sqrt{h^{n}_{j+1}}}{\sqrt{h^{n}_{j}} + \sqrt{h^{n}_{j+1}}};
\end{equation}
and defining $\tilde{J}_\jpc(u^{n}_{j},u^{n}_{j+1})$ as:
\begin{equation}
\tilde{J}_\jpc(u^{n}_{j},u^{n}_{j+1})=\begin{bmatrix}0&1\\c_\text{Roe}^2-v_\text{Roe}^2&2v_\text{Roe}\end{bmatrix};
\end{equation}
The associated eigenvalues $\tilde{\lambda}_\jpc^{(i)}$ and
eigenvectors $\tilde{R}_\jpc^{(i)}$ can easily be computed as:
\begin{equation}\label{eq:Roe_eigenvalues}
\tilde{\lambda}_\jp^{(1)}=v_\text{Roe}-c_\text{Roe}; \quad
\tilde{\lambda}_\jp^{(2)}=v_\text{Roe}+c_\text{Roe};
\end{equation}
\begin{equation}\label{eq:Roe_eigenvectors}
\tilde{R}_\jp^{(1)} = \begin{bmatrix}1\\ \tilde{\lambda}_\jp^{(1)} \end{bmatrix}; \quad
\tilde{R}_\jp^{(2)} = \begin{bmatrix}1\\ \tilde{\lambda}_\jp^{(2)} \end{bmatrix}.
\end{equation}
As usual, the eigenvalues and the eigenvectors are collected in the matrices $\tilde{\Lambda}_\jpc$ and $\tilde{R}_\jpc$.

To introduce the ARoe solution, we integrate \eqref{eq:RP1} in space between $x_j$ and $x_{j+1}$:
\begin{equation}
\frac{\d}{\d t} \int_{x_j}^{x_{j+1}} u(x,t)\, \d x = - \left[f(u(x_{j+1},t)) - f(u(x_j,t))\, \right] + \int_{x_j}^{x_{j+1}} s(x,t)\, \d x;
\end{equation}
and in time between $t^n$ and $t^{n+1} = t^{n} + \Delta t$, taking into account the initial condition \eqref{eq:RP2}:
\begin{equation}
\int_{x_j}^{x_{j+1}} u(x,t^{n+1})\, \d x = \frac{\Delta x (u^{n}_{j}+u^{n}_{j+1})}{2} - \Delta t \left(f^{n}_{j+1} - f^{n}_{j} \right) + \int_{t^n}^{t^{n+1}}\int_{x_j}^{x_{j+1}} s(x,t)\, \d x\, \d t;
\end{equation}
where $f^{n}_{j} = f(u^{n}_{j})$.

Introducing the approximation for the source term integral $\mathcal{S}_\jpc$ given by \cite{Murillo2010,Murillo2013}:
\begin{equation}\label{eq:STint}
\mathcal{S}_\jp = \frac{1}{\Delta t} \int_{t^n}^{t^{n+1}}\int_{x_j}^{x_{j+1}} s(x,t)\, \d x\, \d t =
\begin{bmatrix}
0\\
- \frac{g}{2}\left(h_{j}^{n}+h_{j+1}^{n}\right)\left(z_{j+1} - z_{j}\right)
\end{bmatrix};
\end{equation}
we have:
\begin{equation}\label{eq:intswe}
\int_{x_j}^{x_{j+1}} u(x,t^{n+1})\, \d x = \frac{\Delta x (u^{n}_{j}+u^{n}_{j+1})}{2} - \Delta t \left(f^{n}_{j+1} - f^{n}_{j} \right) + \Delta t\;\mathcal{S}_\jp.
\end{equation}

In the framework of the Roe Riemann solvers, we consider again the RP \eqref{eq:RP1}-\eqref{eq:RP2}, and we write \eqref{eq:RP1} in the following linearized form:
\begin{equation}\label{eq:homogen}
\partial_t \hat{u} + \hat{J}_\jp\partial_x \hat{u} = 0;
\end{equation}
where the Jacobian matrix $\hat{J}_\jpc(u^{n}_{j},u^{n}_{j+1})$  will be specified later. $\hat{J}_\jpc(u^{n}_{j},u^{n}_{j+1})$ includes the effects due to the source term and have to satisfy two conditions: $\hat{J}_\jpc(u^{n}_{j},u^{n}_{j+1})$ must be diagonalizable with real eigenvalues and $\hat{J}_\jpc(u^{n}_{j},u^{n}_{j+1}) \to \hat{J}_\jpc(u^{n}_{j})$ smoothly as $u^{n}_{j+1} \to u^{n}_{j}$.

To provide the explicit form of the matrix $\hat{J}_\jpc$, we perform the following step. We integrate in space and time, obtaining:
\begin{equation}\label{eq:intswelin}
\int_{x_j}^{x_{j+1}} \hat{u}(x,t^n)\, \d x = \frac{\Delta x (u^{n}_{j}+u^{n}_{j+1})}{2} - \Delta t \hat{J}_\jp\left(u^{n}_{j+1} - u^{n}_{j} \right).
\end{equation}
The \emph{consistency condition} (i.e., the integral of the solution $\hat{u}$ is equal to the integral of the solution $u$ over a suitable control volume \cite{Toro09,Murillo2010})  and Eqs.~\eqref{eq:intswe}-\eqref{eq:intswelin} lead to:
\begin{equation}\label{eq:compJFS}
\hat{J}_\jp\left(u^{n}_{j+1} - u^{n}_{j} \right) = \left(f^{n}_{j+1} - f^{n}_{j} \right) - \mathcal{S}_\jp.
\end{equation}
We focus our attention on the left-hand side of Eq.~\eqref{eq:compJFS}, particularly on $\tilde{J}_\jpc$ and the corresponding eigenvectors and eigenvalues. Then, we can write:
\begin{equation}
\tilde{J}_\jp = \tilde{R}_\jp \tilde{\Lambda}_\jp \tilde{R}_\jp^{-1}.
\end{equation}
We consider the projection of the difference $u^{n}_{j+1} - u^{n}_{j}$ and the source term $\mathcal{S}_\jpc$ over the orthogonal basis constituted by the eigenvectors of the matrix $\tilde{R}_\jpc$, i.e., we compute the coefficients $\alpha_\jpc=[\alpha_\jpc^{(1)},\alpha_\jpc^{(2)}]^{\text{T}}$ and $\beta_\jpc=[\beta_\jpc^{(1)},\beta_\jpc^{(2)}]^{\text{T}}$ that satisfy:
\begin{equation}\label{eq:alpha_beta}
\tilde{R}_\jp \, \alpha_\jp= u^{n}_{j+1} - u^{n}_{j}; \qquad \tilde{R}_\jp\,\beta_\jp=\mathcal{S}_\jp.
\end{equation}
and we can apply to the right-hand side of the Eq.~\eqref{eq:compJFS} the following manipulation:
\begin{multline} 
\left(f^{n}_{j+1} - f^{n}_{j} \right) - \mathcal{S}_\jp = \tilde{J}_\jp\left(u^{n}_{j+1} - u^{n}_{j} \right) - \mathcal{S}_\jp=\\
= \tilde{R}_\jp \tilde{\Lambda}_\jp \tilde{R}_\jp^{-1}\left(u^{n}_{j+1} - u^{n}_{j} \right) - \mathcal{S}_\jp=\\
= \tilde{R}_\jp \tilde{\Lambda}_\jp \tilde{R}_\jp^{-1}\tilde{R}_\jp \, \alpha_\jp - \tilde{R}_\jp\,\beta_\jp =\\
= \tilde{R}_\jp \left(\tilde{\Lambda}_\jp\alpha_\jp - \beta_\jp \right)
= \sum_{i=1}^2 \left(\tilde{\lambda}_\jp^{(i)} \alpha_\jp^{(i)} - \beta_\jp^{(i)} \right) \tilde{R}_\jp^{(i)};
\end{multline}
or in a more compact form:
\begin{equation}\label{eq:RHS}
\left(f^{n}_{j+1} - f^{n}_{j} \right) - \mathcal{S}_\jp = 
\sum_{i=1}^2 \left[\tilde{\lambda}\, \alpha - \beta \right]_\jp^{(i)} \tilde{R}_\jp^{(i)};
\end{equation}

Introducing the relationship:
\begin{equation}\label{eq:scomp_hatJ}
\hat{J}_\jp = \tilde{R}_\jp \hat{\Lambda}_\jp \tilde{R}_\jp^{-1};
\end{equation}
where $\hat{\Lambda}$ is the diagonal matrix of eigenvalues $\hat{\lambda}_i$, the left-hand side of the Eq.~\eqref{eq:compJFS} can be manipulated in the following manner:
\begin{multline}
\hat{J}_\jp\left(u^{n}_{j+1} - u^{n}_{j} \right) = \tilde{R}_\jp \hat{\Lambda}_\jp \tilde{R}_\jp^{-1}\left(u^{n}_{j+1} - u^{n}_{j} \right) =\\
= \tilde{R}_\jp \hat{\Lambda}_\jp \tilde{R}_\jp^{-1}\tilde{R}_\jp \, \alpha_\jp 
  = \tilde{R}_\jp \left(\hat{\Lambda}_\jp\alpha_\jp\right) =\\
  = \sum_{i=1}^2 \left(\hat{\lambda}_\jp^{(i)} \alpha_\jp^{(i)} \right) \tilde{R}_\jp^{(i)}.
\end{multline}
or in a more compact form:
\begin{equation}\label{eq:LHS}
\hat{J}_\jp\left(u^{n}_{j+1} - u^{n}_{j} \right) 
= \sum_{i=1}^2 \left[\hat{\lambda} \alpha \right]_\jp^{(i)} \tilde{R}_\jp^{(i)}.
\end{equation}

Taking into account Eqs.~\eqref{eq:compJFS}, \eqref{eq:RHS} and \eqref{eq:LHS}, we have:
\begin{equation}
\sum_{i=1}^2 \left[\tilde{\lambda}\, \alpha - \beta \right]_\jp^{(i)} \tilde{R}_\jp^{(i)}
=\sum_{i=1}^2 \left[\hat{\lambda} \alpha \right]_\jp^{(i)} \tilde{R}_\jp^{(i)};
\end{equation}
which allows defining the eigenvalues $\hat{\lambda}_i$ as:
\begin{equation}\label{eq:hatlambda}
\hat{\lambda}_\jp^{(i)} = \tilde{\lambda}_\jp^{(i)} - \frac{\beta_\jp^{(i)}}{\alpha_\jp^{(i)}};
\end{equation}

Given a suitable Roe matrix for the homogeneous part of the system, $\tilde{J}_\jpc$, the construction of $\hat{J}_\jpc$ is performed using \eqref{eq:hatlambda} and \eqref{eq:scomp_hatJ}.

In the framework of a Roe scheme, an approximate solution of the RP
associated with \eqref{eq:homogen} can easily be found. In the classical approach, the approximate solution consists of piecewise constant functions with the number of discontinuities equal to the number of equations in the hyperbolic system. In our case, the classical solutions are constituted by three constant states separated by two waves. In the approach proposed in \cite{Murillo2010}, a new state and a new wave are introduced. Whereas the original waves are characterized by a celerity related to the Jacobian matrix eigenvalues, the new wave is steady. The corresponding \emph{three-wave solution} \cite{Murillo2010} depends on the flow condition, that is, on the signs of $\tilde{\lambda}_i$, and takes into account the presence of the bottom discontinuity.

In the case of subcritical flows, the solution is given by \cite{Murillo2010}:
\begin{equation}\label{eq:subsol}
\hat{u}(x,t)=\left\{\begin{array}{ll}
u_{j}^{n}, &\text{if}\ (x-x_\jp) \leq \tilde{\lambda}_1 t;\\
u_{j}^{*}, &\text{if}\ (x-x_\jp) > \tilde{\lambda}_1 t\ \text{and}\ x \leq x_\jp; \\
u_{j+1}^{**}, &\text{if}\ x > x_\jp\ \text{and}\ (x-x_\jp) < \tilde{\lambda}_2 t; \\
u_{j+1}^{n}, &\text{if}\ (x-x_\jp) \geq \tilde{\lambda}_2 t; \end{array} \right.
\end{equation}
with:
\begin{align}
u_{j}^{*} &= u_{j}^{n} + \left[\alpha - \frac{\beta}{\tilde{\lambda}} \right]_\jp^{(1)} \tilde{R}_\jp^{(1)};\\
u_{j+1}^{**} & = u_{j+1}^{*} - \left[\alpha - \frac{\beta}{\tilde{\lambda}} \right]_\jp^{(2)} \tilde{R}_\jp^{(2)};
\end{align}

In the supercritical case, if $v>0$, the solution becomes:
\begin{equation}
\hat{u}(x,t)=\left\{\begin{array}{ll}
u_{j}^{n}, &\text{if}\ x \leq x_\jp;\\
u_{j+1}^{*}, &\text{if}\ x > x_\jp\ \text{and},\ (x-x_\jp) \leq x \tilde{\lambda}_1 t; \\
u_{j+1}^{**}, &\text{if}\ (x-x_\jp) > \tilde{\lambda}_1 t\ \text{and}\ (x-x_\jp) < \tilde{\lambda}_2 t, \\
u_{j+1}^{n}, &\text{if}\ (x-x_\jp) \geq \tilde{\lambda}_2 t; \end{array} \right.
\end{equation}
with:
\begin{equation}
u_{j+1}^{*} = u_{j+1}^{**} - \left[\alpha - \frac{\beta}{\tilde{\lambda}} \right]_\jp^{(1)} \tilde{R}_\jp^{(1)};
\end{equation}

In the supercritical case, if $v<0$, the solution becomes:
\begin{equation}
\hat{u}(x,t)=\left\{\begin{array}{ll}
u_{j}^{n}, &\text{if}\ (x-x_\jp) \leq \tilde{\lambda}_1 t;\\
u_{j}^{*}, &\text{if}\ (x-x_\jp) > \tilde{\lambda}_1 t\ \text{and}\ (x-x_\jp) < \tilde{\lambda}_2 t; \\
u_{j}^{**}, &\text{if}\ (x-x_\jp) > \tilde{\lambda}_2 t\ \text{and}\ x < x_\jp; \\
u_{j+1}^{n}, &\text{if}\ x \geq x_\jp; \end{array} \right.
\end{equation}
with:
\begin{equation}
u_{j}^{**} = u_{j}^{*} + \left[\alpha - \frac{\beta}{\tilde{\lambda}} \right]_\jp^{(2)} \tilde{R}_\jp^{(2)};
\end{equation}

With this approximate solution, we are now able to write the expression for updating the cell-averaged state. To this end, we can integrate the approximate solutions of the two Riemann problems at the cell interfaces as usual in the Godunov-type schemes.
Without loss of generality, we consider the subcritical case and the cell $I_j = \left[x_\jmc,x_\jpc\right]$. The cell-averaged values $u_j^{n+1}$ on the $j$-th cell at time $t^{n+1}$ are defined as:
\begin{equation}\label{eq:updatingARoe}
u_j^{n+1} = \frac{1}{\Delta x} \int_{I_j} u(x,t^{n+1})\, \d x \approx \frac{1}{\Delta x} \int_{I_j} \hat{u}(x,t^{n+1})\, \d x;
\end{equation}

The substitution of the approximate solution \eqref{eq:subsol} in \eqref{eq:updatingARoe} allows writing:
\begin{equation}
u_j^{n+1} \Delta x =
u_j^{*}\left(-\tilde{\lambda}_\jp^{(1)} \Delta t\right)
+ u_j^{n}\left(\Delta x 
- \tilde{\lambda}_\jm^{(2)} \Delta t
+ \tilde{\lambda}_\jp^{(1)} \Delta t\right)
+ u_j^{**}\left(\tilde{\lambda}_\jm^{(2)} \Delta t\right);
\end{equation}
that can be written as:
\begin{equation}
u_j^{n+1} \Delta x =  u_j^{n} \Delta x
+ \left(u_j^{n} - u_j^{*}\right)\left(\tilde{\lambda}_\jp^{(1)} \Delta t\right)
+ \left(u_j^{**}- u_j^{n}\right)\left(\tilde{\lambda}_\jm^{(2)} \Delta t\right);
\end{equation}
and straightforward algebraic manipulations lead to:
\begin{equation}
u_j^{n+1} =  u_j^{n} - \frac{\Delta t}{\Delta x} \left\{
\left[\left(\tilde{\lambda}\alpha - \beta \right) \tilde{R}\right]_\jp^{(1)}
+ \left[\left(\tilde{\lambda}\alpha - \beta \right) \tilde{R}\right]_\jm^{(2)} \right\}.
\end{equation}

Analogous computations that take into account both subcritical and supercritical flows provide the general updating expression for the numerical solution:
%
\begin{equation} \label{eq:updateARoe}
u_j^{n+1} = u_j^{n} - \frac{\Delta t}{\Delta x} \left[\tilde{\mathcal{D}}^{-}_{\jp} + \tilde{\mathcal{D}}^{+}_{\jm} \right];
\end{equation} 
with:
\begin{align}
\tilde{\mathcal{D}}^{-}_{\jp} &= \sum_{\tilde{\lambda}_\jp^{(i)}<0}
\left[\left(\tilde{\lambda}\alpha - \beta \right) \tilde{R}\right]_\jp^{(i)}; \label{eq:fltARoe1}\\
\tilde{\mathcal{D}}^{+}_{\jm} &= \sum_{\tilde{\lambda}_\jm^{(i)}>0}
\left[\left(\tilde{\lambda}\alpha - \beta \right) \tilde{R}\right]_\jm^{(i)}; \label{eq:fltARoe2}
\end{align}
where the sum in \eqref{eq:fltARoe1} is performed on the indices $i$ for which $\tilde{\lambda}_\jp^{(i)}<0$, while the sum in \eqref{eq:fltARoe2} is performed on the indices $i$ for which $\tilde{\lambda}_\jm^{(i)}>0$.

To avoid entropy glitches \cite{Toro09} in the case of transcritical flows, an Harten-Hyman entropy fix is used; see \cite{Murillo2010} for further details.

The ARoe scheme can be summarized by the following steps: the source
term integral $\mathcal{S}_\jpc$ is computed using
Eq.~\eqref{eq:STint}; the eigenvalues $\tilde{\lambda}_\jpc^{(i)}$ and
eigenvectors $\tilde{R}_\jpc^{(i)}$ are computed using
\eqref{eq:Roe_eigenvalues} and \eqref{eq:Roe_eigenvectors},
respectively; $\alpha_\jpc$ and $\beta_\jpc$ are computed using
\eqref{eq:alpha_beta}; $\tilde{\mathcal{D}}^{-}_{\jpc}$ and
$\tilde{\mathcal{D}}^{+}_{\jmc}$ are evaluated using
\eqref{eq:fltARoe1} and \eqref{eq:fltARoe2}, respectively; and finally, the numerical solution is updated through \eqref{eq:updateARoe}.
%
\subsection{The ARoeEB model - an energy-balanced augmented Roe model} \label{sec:ARoeWB}
The ARoe model described in \S~\ref{sec:ARoe} is well balanced for the quiescent flow \cite{Murillo2010}.
To obtain an energy-balanced model \cite{Murillo2013}, denoted here as
ARoeEB, the integral $\mathcal{S}_\jpc$:
\begin{equation}\label{eq:Scomponents}
\mathcal{S}_\jp = \frac{1}{\Delta t} \int_{t^n}^{t^{n+1}}\int_{x_j}^{x_{j+1}} s(x,t)\, \d x\, \d t =
\begin{bmatrix}
0 \\
\mathcal{S}_\jp^{(2)}
\end{bmatrix};
\end{equation}
must be modified. First, we recall that $\mathcal{S}_\jpc^{(2)}$ can be considered as the force exerted by a step on the flow \cite{Caleffi2009a,Murillo2013,Caleffi2015a} and can be evaluated in different manners. 

If we assume that the pressure distribution is hydrostatic over the step and depends only on the free-surface level on the side of the discontinuity where the bottom elevation is lower \cite{Murillo2013}, we obtain the $s_{b1}$ approximation given by:
\begin{equation}\label{eq:sb1}
s_{b1} = - g\left(h_{l}-\frac{|\jump{z}'|}{2}\right)\jump{z}';
\end{equation}
where: 
\begin{equation}
l=\left\{\begin{array}{ll}
  j, &\text{if}\ \jump{z} \geq 0;\\
j+1, &\text{if}\ \jump{z} < 0;
\end{array} \right. \qquad
\jump{z}'=\left\{\begin{array}{ll}
h_{j}^{n}, &\text{if}\ \jump{z} \geq 0 \text{ and } \eta_{j}^{n}<z_{j+1};\\
h_{j+1}^{n}, &\text{if}\ \jump{z} < 0 \text{ and } \eta_{j+1}^{n}<z_{j};\\
\jump{z}, &\text{otherwise};
\end{array} \right.
\end{equation}
with:
$\jump{z} = z_{j+1} - z_{j}$ and $\eta_{j}^{n} = h_{j}^{n} + z_{j}$.

A simpler approximation, denoted here as $s_{b2}$, is given by:
\begin{equation}\label{eq:sb2}
s_{b2} = - g\bar{h}_\jp^{n}\jump{z};
\end{equation}
with $\bar{h}_\jp^{n}= \left(h_{j}^{n}+h_{j+1}^{n}\right)/2$.

The key idea is to combine Eqs.~\eqref{eq:sb1}-\eqref{eq:sb2} to obtain an approximation of $\mathcal{S}_\jp^{(2)}$ that allows defining an energy-balanced scheme. To this end, the following linear combination is proposed in \cite{Murillo2013}:
\begin{equation}\label{eq:lincombS}
\mathcal{S}_\jp^{(2)} = \left(1-\mathcal{A}\right) s_{b2} + \mathcal{A}\, s_{b1};
\end{equation}
with $\mathcal{A}$ to be defined.

To define the weight $\mathcal{A}$, the first step consists of
determining the value that $\mathcal{A}$ must assume in steady
conditions without shocks. We denote this value as $\mathcal{A}_E$.
In this case, the total head must be conserved across the bottom discontinuity, and therefore, we can write:
\begin{equation}\label{eq:hcost1}
\jump{z+h+v^2/(2\,g)} = 0.
\end{equation}
Eq.~\eqref{eq:hcost1} can be cast in the form:
\begin{equation}\label{eq:hcost2}
\jump{z+h} = -\frac{1}{g}\jump{v^2/2}.
\end{equation}

For the same steady condition, straightforward computations performed on Eq.~\eqref{eq:intswe} provide:
\begin{align}
\jump{q} &= 0;\\
\jump{h\,v^2 + (g\,h^2)/2} &= \left(1-\mathcal{A}_E\right) s_{b2} + \mathcal{A}_E\, s_{b1};\label{eq:QdMARoeEB}
\end{align}
and \eqref{eq:QdMARoeEB} becomes:
\begin{equation}\label{eq:QdMARoeEB2}
\jump{h\,v^2} + g \bar{h}_\jp^{n} \jump{h} - s_{b2} = \mathcal{A}_E\left(s_{b1}-s_{b2}\right);
\end{equation}
being $\jump{(g\,h^2)/2}=g\bar{h}_\jp^{n} \jump{h}$. The substitution of \eqref{eq:sb2} into \eqref{eq:QdMARoeEB2} leads to:
\begin{equation}\label{eq:QdMARoeEB3}
\jump{h\,v^2} + g \bar{h}_\jp^{n} \jump{z+h} = \mathcal{A}_E\left(s_{b1}-s_{b2}\right).
\end{equation}

If we impose that the momentum balance and the total head preservation have to be valid at the same time when the flow is steady, both \eqref{eq:hcost2} and \eqref{eq:QdMARoeEB3} must be true, and therefore:
\begin{equation}\label{eq:QdMARoeEB4}
\jump{h\,v^2} - \bar{h}_\jp^{n} \jump{v^2/2} = \mathcal{A}_E\left(s_{b1}-s_{b2}\right);
\end{equation}
which implies that:
\begin{equation}\label{eq:AE}
\mathcal{A}_E = \frac{\jump{h\,v^2} - \bar{h}_\jp^{n} \jump{v^2/2}}{s_{b1}-s_{b2}}.
\end{equation}

Eq.~\eqref{eq:AE} ensures that, in the case of steady conditions and
away from hydraulic jumps, the total head is preserved if $\mathcal{A}
= \mathcal{A}_E$. Conversely, in the presence of a hydraulic jump, a proper fraction of the total head must be dissipated. Therefore, a robust algorithm to compute both steady and unsteady flows, with or without hydraulic jumps, is necessary.
As suggested in \cite{Murillo2013}, the following solution is used in this work:
\begin{equation}\label{eq:defA}
\mathcal{A}=\left\{\begin{array}{ll}
\mathcal{A}_E, &\text{if } v_{j}^{n}v_{j+1}^{n} \geq 0 \text{ and } |\ten{Fr}_j|<1 \text{ and } |\ten{Fr}_{j+1}|<1;\\
\mathcal{A}_E, &\text{if } v_{j}^{n}v_{j+1}^{n} \geq 0 \text{ and } |\ten{Fr}_j|>1 \text{ and } |\ten{Fr}_{j+1}|>1;\\
1, &\text{otherwise}.
\end{array} \right.
\end{equation}
This algorithm discriminates between smooth and discontinuous
solutions \cite{Murillo2013}. In the case of steady flows without
hydraulic jumps, the solution guarantees the total head preservation,
while the still water case is correctly captured as a particular case
of steady flow. In the presence of a hydraulic jump, by imposing  $\mathcal{A} = 1$, the correct rate of energy dissipation is obtained.

For completeness, we summarize the ARoeEB scheme here: the source term
integral $\mathcal{S}_\jpc$ is computed using
Eqs.~\eqref{eq:Scomponents} and \eqref{eq:lincombS} with $\mathcal{A}$
given by Eq.~\eqref{eq:defA}; the eigenvalues
$\tilde{\lambda}_\jpc^{(i)}$ and eigenvectors $\tilde{R}_\jpc^{(i)}$
are computed using \eqref{eq:Roe_eigenvalues} and
\eqref{eq:Roe_eigenvectors}, respectively; $\alpha_\jpc$ and
$\beta_\jpc$ are computed using \eqref{eq:alpha_beta};
$\tilde{\mathcal{D}}^{-}_{\jpc}$ and $\tilde{\mathcal{D}}^{+}_{\jmc}$
are evaluated using \eqref{eq:fltARoe1} and \eqref{eq:fltARoe2},
respectively; and finally, the numerical solution is updated through \eqref{eq:updateARoe}. 
\subsection{The HDEB model - a hydrodynamic reconstruction model} \label{sec:HDR}
Here, we also consider the hydrodynamic reconstruction model, denoted HDEB, proposed in \cite{Caleffi2009a} and revisited in \cite{Caleffi2015a}. 

The scheme is developed assuming the conservation of the total head and of the specific discharge on the step for the steady condition. The scheme belongs to the family of the finite volume methods, and the expression for the updating of the solution is:
\begin{equation} \label{eq:updateHDR}
u_j^{n+1} = u_j^{n} - \frac{\Delta t}{\Delta x} \left[ f^{-}_{\jp} - f^{+}_{\jm} \right];
\end{equation} 
where $f^{-}_{\jpc}$ and $f^{+}_{\jmc}$ are suitable numerical fluxes to be given.

Recalling that the total force, $\Phi(u)$, and the specific energy, $E(u)$, are given by:
\begin{equation}
\Phi(u) = \frac{g\,h^2}{2} + \frac{q^2}{h}; \qquad E(u) = h + \frac{q^2}{2\,g\,h^2};
\end{equation}
the numerical fluxes $f^{-}_{\jpc}$ and $f^{+}_{\jmc}$ are:
\begin{align}
f^{-}_{\jp} &= f^{*}(u^{*,-}_\jp,u^{*,+}_\jp) + 
\begin{bmatrix} 0\\ \Phi\left(u^{n}_{j}\right) - \Phi\left(u^{*,-}_\jp\right)
\end{bmatrix};  \label{eq:HDRflux1}\\
f^{+}_{\jm} &= f^{*}(u^{*,-}_\jm,u^{*,+}_\jm) + 
\begin{bmatrix} 0\\ \Phi\left(u^{n}_{j}\right) - \Phi\left(u^{*,+}_\jm\right)
\end{bmatrix}; \label{eq:HDRflux2}
\end{align}
where $f^*(u_L,u_R)$ is the standard HLL numerical flux \cite{HLL,Toro09} and $u^{*,\pm}_{j \pm 1}$ is defined as follows. The attention is focused on $u^{*,-}_\jpc$, and a similar procedure can be constructed for $u^{*,+}_\jmc$.

A virtual section between the $j$-th and $j+1$-th cells is introduced, and a layer of infinitely small length between the interface at $x_\jpc$ of the $j$-th cell and the virtual section is considered.
We assume that the bottom elevation at the virtual section can be computed as $z_\jp^*=\max(z_{j},z_{j+1})$.
Then, we compute:
\begin{equation}
E_\jp^{*,-}=(z_{j} - z_\jp^{*}) + E_{j}^{n};
\end{equation}
with $E_{j}^{n} = E(u_{j}^{n})$. This relation is obtained by imposing the conservation of the total head and of the discharge into the virtual layer. Then, $E_\jp^{*,-}$ can be interpreted as the function $E$ computed at the virtual section at $x_\jp^-$, i.e.:
\begin{equation}\label{eq:espi}
E_\jp^{*,-} = h_\jp^{*,-} + \frac{(q_{j}^{n})^2}{2\,g(h_\jp^{*,-})^2}
\end{equation}
that corresponds to an implicit expression for $h_\jp^{*,-}$. Finding
$h_\jp^{*,-}$ that satisfies Eq.~\eqref{eq:espi} for given values of
$q_{j}^{n}$ and $E_\jp^{*,-}$ is performed analytically
\cite{Valiani2008}. Two depths make physical sense and correspond to a
subcritical solution and a supercritical solution. In this work, we select the solution consistent with the Froude number $\mathsf{Fr}(u_{j}^{n})$.

The HDEB scheme can be summarized in the following steps: we compute
$u^{*,-}_\jp=[h_\jp^{*,-},q_{j}^{n}]^\text{T}$, where $h_\jp^{*,-}$ is
given implicitly by \eqref{eq:espi}; we compute the HLL numerical
fluxes and the corrections; the HDEB fluxes are then computed using
Eqs.~\eqref{eq:HDRflux1}-\eqref{eq:HDRflux2}; and finally, the solution is updated using \eqref{eq:updateHDR}.

\section{Comparison between models}\label{sec:comparison}
Several test cases are used to compare the different models. Only the
meaningful results are presented here. The CFL coefficient is set
equal to 0.9 in almost all the simulations with all the models, but in
some particular cases, the time step is decreased to achieve
meaningful results. In these specific cases, the CFL used is specified in the description of the tests.
%
\subsection{C-property test case}\label{sec:cproperty}
To verify the fulfillment of the C-property over a non-flat bottom \cite{BeVa-94}, the test case proposed in \cite{Caleffi2015a} is used. The bottom profile is continuous and differentiable but with two discontinuities  and is given by:
\begin{equation}
z(x)=\left\{\begin{array}{lll}
\sin(2\pi x) & \text{if} & 0.0 \text{ m} < x \leq 0.4 \text{ m};\\
\cos(2\pi (x+1)) & \text{if} & 0.4 \text{ m} < x < 0.8 \text{ m};\\
\sin(2\pi x) & \text{if} & 0.8 \text{ m} < x \leq 1.0 \text{ m};
\end{array}\right.
\end{equation}
The initial conditions are a constant free-surface elevation, $\eta$ = 1.5 m, and a zero discharge, whereas the boundary conditions are periodic. The simulations are performed until $t$ = 0.1 s, using a mesh of 20 cells.
The analytical solution consists of a quiescent flow that preserves the initial free-surface elevation.

The $L_1$, $L_2$ and $L_{\infty}$ error norms related to the water level and the specific discharge are computed. The results, obtained using double-precision floating-point arithmetic, are summarized in Tab.~\ref{tab:C-property}. 
\begin{table} 
\centering
\begin{small}
\begin{tabular}{l c c c | c c c} 
\hline 
\hline 
       & \multicolumn{3}{c|}{$\eta$} & \multicolumn{3}{c}{$q$}\\
\hline 
Model & $L_1$ & $L_2$ & $L_{\infty}$ & $L_1$ & $L_2$ & $L_{\infty}$\\ 
\hline \hline 
PC       & 4.441e-17 & 1.404e-16 & 4.441e-16 & 3.592e-16 &  4.698e-16 & 9.557e-16 \\
ARoe     & 9.992e-17 & 2.047e-16 & 4.441e-16 & 3.543e-16 &  4.259e-16 & 7.274e-16 \\
HLLEM    & 5.551e-17 & 1.440e-16 & 4.441e-16 & 4.911e-16 &  6.230e-16 & 1.531e-15 \\
PCEB     & 2.220e-17 & 9.930e-17 & 4.441e-16 & 6.119e-17 &  1.212e-16 & 3.296e-16 \\
ARoeEB   & 9.992e-17 & 2.047e-16 & 4.441e-16 & 3.543e-16 &  4.259e-16 & 7.274e-16 \\
HLLEMEB  & 2.220e-17 & 9.930e-17 & 4.441e-16 & 2.755e-16 &  3.780e-16 & 8.511e-16 \\
HDEB     & 0.000e+00 & 0.000e+00 & 0.000e+00 & 2.808e-17 &  8.432e-17 & 3.186e-16 \\
\hline 
\hline 
\end{tabular}
\end{small}
\caption{C-property test case: the $L_1$, $L_2$ and $L_{\infty}$ error norms in terms of water elevation $\eta$ and specific discharge $q$, for a motionless steady flow, are shown for all the models.} \label{tab:C-property} 
\end{table} 

The differences of the numerical solutions from the reference solution
are only due to round-off errors; therefore, all the models considered
in this work, both the well-balanced and the energy-balanced models,
are able to exactly satisfy the C-property.

\subsection{Subcritical and supercritical flows over a bump}\label{sec:subsuper}
The purpose of this test case is to validate the seven models in the
case of steady flows over a parabolic bottom profile
\cite{Valiani2002a,Caleffi2003,Caleffi2015a}. In particular, we are
interested in comparing between the energy-balanced models
\cite{Murillo2013} and the models that are well balanced only for the
quiescent flow \cite{BeVa-94}. In this section, we take flows without
a transition through the critical state into account.

We consider a 20 m long channel, discretized with 400 cells ($\Delta x = 0.05$ m). The bottom profile is given by the following function:
\begin{equation}
z(x)=\left\{\begin{array}{ll}
0.2 - 0.05\left(x-10\right)^2 \text{ m}\qquad & \text{if } 8.0 \text{ m} < x \leq 12 \text{ m};\\
0 & \text{otherwise};
\end{array}\right.
\end{equation}
We consider both subcritical and supercritical flows. In the case of subcritical flow, the upstream specific discharge is imposed equal to 4.42 m$^2$/s, while the downstream water level is imposed equal to 2.00 m. As initial conditions, we set a uniform specific discharge and a uniform depth of 4.42 m$^2$/s and 2.00 m, respectively. In the case of supercritical flow, we impose at the upstream section a specific discharge and a depth of 4.42 m$^2$/s and 0.85 m, respectively. The initial conditions are again uniform with specific discharge and depth equal to the corresponding upstream boundary values.

\begin{figure} 
\begin{center}
\includegraphics[width=\textwidth]{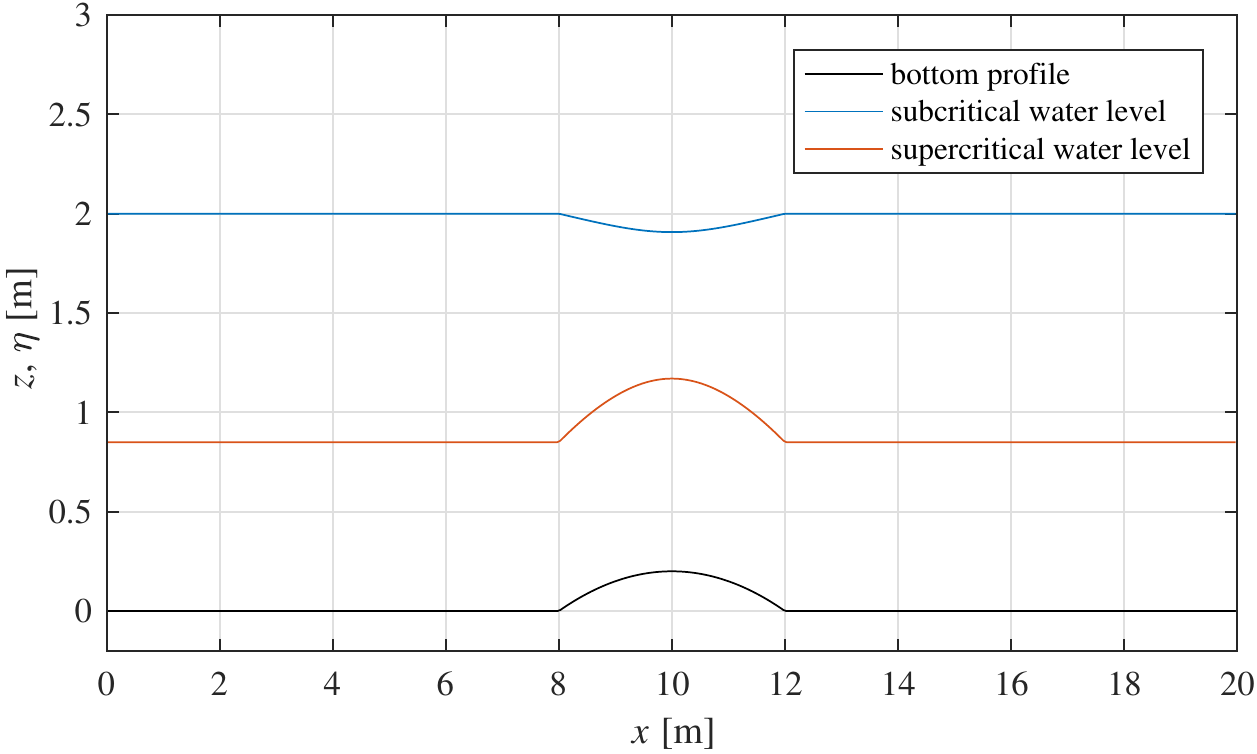}
\end{center}
\caption{Bottom profile and free-surface water level for the subcritical and supercritical flows over a bump.}\label{fig:sssketch}
\end{figure}

After an initial transitory flow, the constant boundary conditions
must lead to an asymptotic steady state characterized by a uniform specific discharge and total head. The corresponding asymptotic analytical solutions are reported in Fig.~\ref{fig:sssketch} in terms of free surface elevation for both the subcritical and supercritical states.  We assume that the steady state is reached after 100 s of simulation.

Note that we can define a scheme as energy balanced if, in the absence
of a hydraulic jump, it can exactly preserve an initial steady state
characterized by a uniform discharge and total head \cite{Murillo2013}. The initial conditions that we have selected do not satisfy the total head uniformity, and the models must reproduce an unsteady flow toward the steady asymptotic solution. Therefore, a numerical model works correctly only if the correct asymptotic state is reached and maintained indefinitely. Our choice of the initial conditions makes the test more demanding with respect to the simple steady state preservation because the asymptotic state is not imposed by the initial condition but is computed by the models.

The analysis is presented in terms of relative errors $e_q$ and $e_H$ in the computation of the asymptotic specific discharge and total head defined as:
\begin{equation} \label{eq:error_def}
e_q=\frac{q-q_{\text{ref}}}{q_{\text{ref}}}; \qquad e_H=\frac{H-H_{\text{ref}}}{H_{\text{ref}}};
\end{equation}
where $q_{\text{ref}}$ and $H_{\text{ref}}$ are the specific discharge
and the total head of the reference analytical solution, respectively.

\begin{figure} 
\begin{center}
\includegraphics[width=0.92\textwidth]{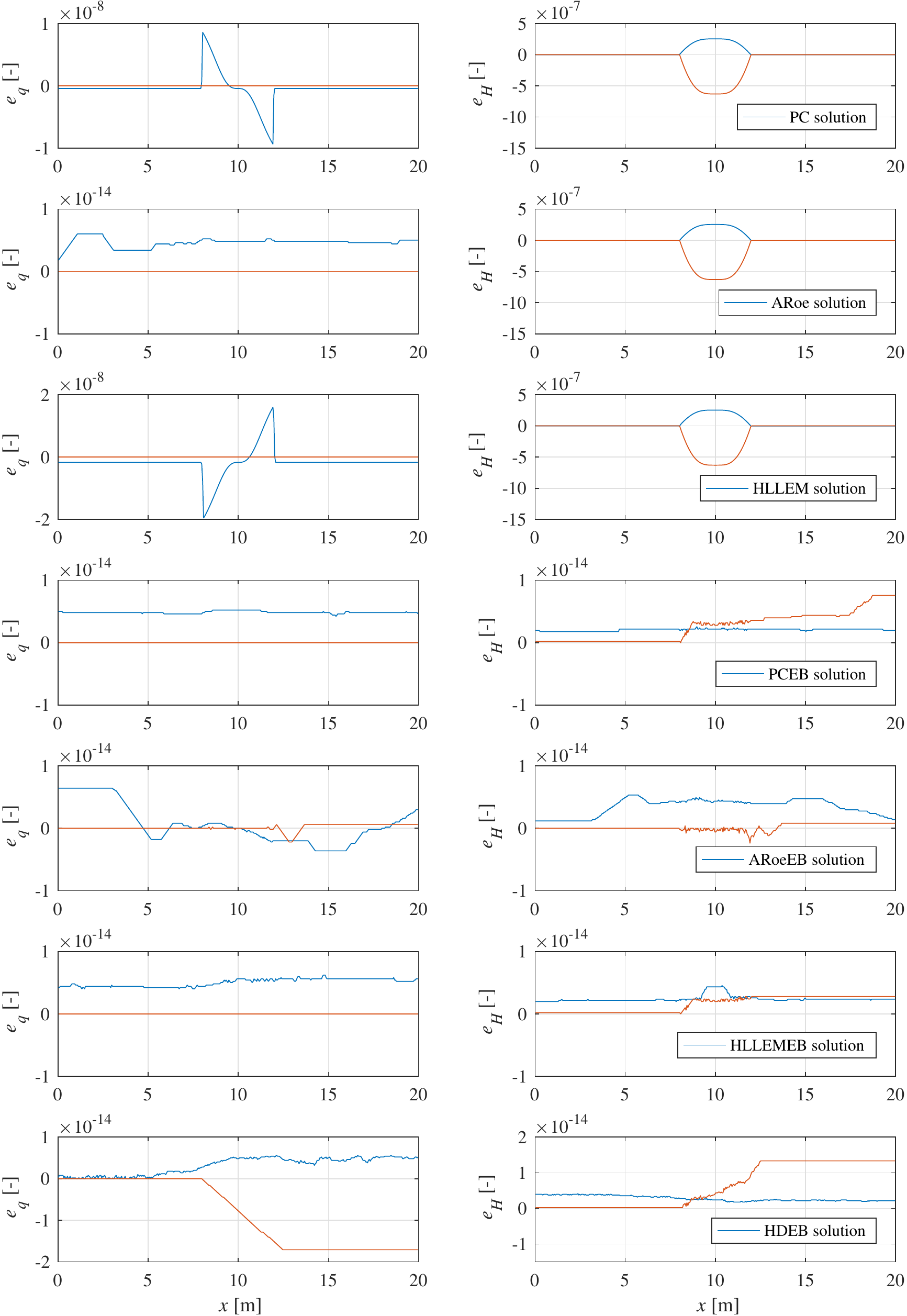}
\end{center}
\caption{Subcritical and supercritical flows: comparison between
  errors in the flow discharge and in the total head. Blue lines refer
  to the subcritical flow, and red lines refer to the supercritical flow.}\label{fig:subsuper}
\end{figure}

In Fig.~\ref{fig:subsuper}, $e_q$ and $e_H$ are reproduced for all the considered models for the subcritical case (blue lines) and for the supercritical case (red lines).

As expected, two different trends can be observed in the
results. Whereas the well-balanced models (PC, ARoe and HLLEM) are not
able to reproduce the steady state exactly (at the round-off error
level), all the four energy-balanced models (PCEB, ARoeEB, and
HLLEMEB) reach the asymptotic solutions exactly. This can clearly be
observed when examining the figures of the total head error. In
particular, the analysis of the results obtained using the WB models
shows a maximum relative error in the total head computation on the
order of $10^{-7}$, and interestingly, all three models show exactly
the same behavior. Moreover, the sign of the error is equal for all
the EB models in the case of both subcritical and supercritical flows. Moreover, note that in our experience with the ARoeEB model, the related time step must be reduced to achieve acceptable results for the supercritical test case (i.e., the CFL coefficient is 0.5).

Small differences can be observed in the reproduction of the flow
discharge. For example, while the PC and HLLEM models do not allow
achieving a uniform exact specific discharge, as expected by the WB models, the ARoe model provides exact results. Moreover, the performances of the models in the reproduction of the supercritical flow appear to be better than in the reproduction of the subcritical flows.

The simulation of the subcritical flow over the bump is also used to
study the numerical efficiency of the different models. To the best of
our abilities, all the models are developed with the same degree of
optimization. In Tab.~\ref{tab:efficiency}, the ratios between the
simulation time of each model and the simulation time of the HLLEM
model are shown. The HLLEM model results in the more efficient model,
whereas the longest time is achieved using the PCEB model. In general,
the techniques that preserve the total head lead to a certain increase
in the computational time.
\begin{table} 
\centering
\begin{small}
\begin{tabular}{c c c c c c c} 
\hline \hline 
PC & ARoe & HLLEM & PCEB & ARoeEB & HLLEMEB  & HDEB \\
\hline
1.04 &   1.29 &   1.00 &  13.49 &   3.29  &  2.84 &   4.47 \\
\hline \hline 
\end{tabular}
\end{small}
\caption{Relative simulation time for the subcritical test case: ratio between the simulation time of each model and the HLLEM model.} \label{tab:efficiency} 
\end{table} 
%
%
\subsection{Transcritical flow over a bump}\label{sec:transbump}
The total head preservation becomes more difficult to achieve if the solution involves critical states or hydraulic jumps. In this test case, we consider a smooth transcritical flow over a parabolic bump.

\begin{figure} 
\begin{center}
\includegraphics[width=\textwidth]{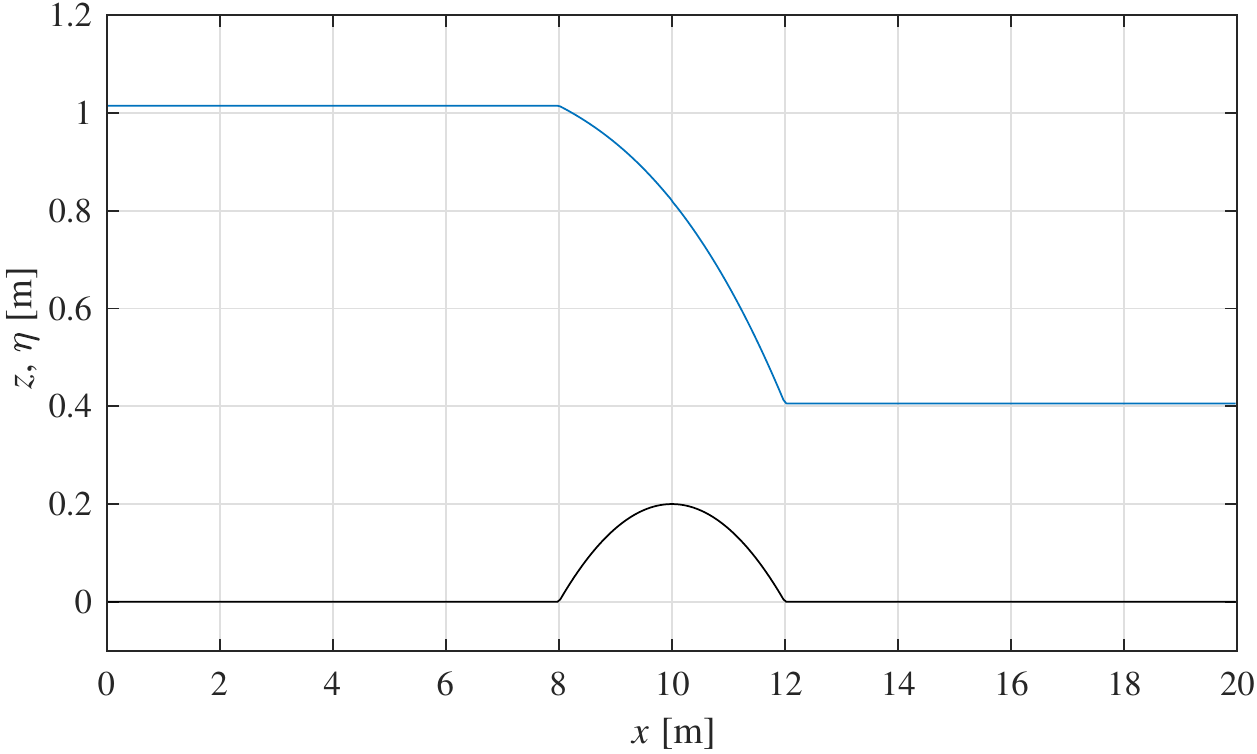}
\end{center}
\caption{Bottom profile and free-surface water level for the transcritical flow over the bump.}\label{fig:transbump}
\end{figure}

We consider the same channel, bottom profile and space discretization
of the test case presented in section \ref{sec:subsuper}. The boundary
conditions are imposed to obtain the transition through the critical
state on the vertex of the bump, a subcritical flow in the upstream
part of the channel and a supercritical flow in the downstream part. To
this end, at the upstream section, the specific discharge is set to 1.53 m$^2$/s. A uniform specific discharge and a uniform water level of 0.4 m are initially imposed along the channel. After the initial transient, the asymptotic steady state is reached after 100 s of simulation. The analytical solution is reported in Fig.~\ref{fig:transbump} in terms of free surface elevation.

\begin{figure} 
\begin{center}
\includegraphics[width=0.92\textwidth]{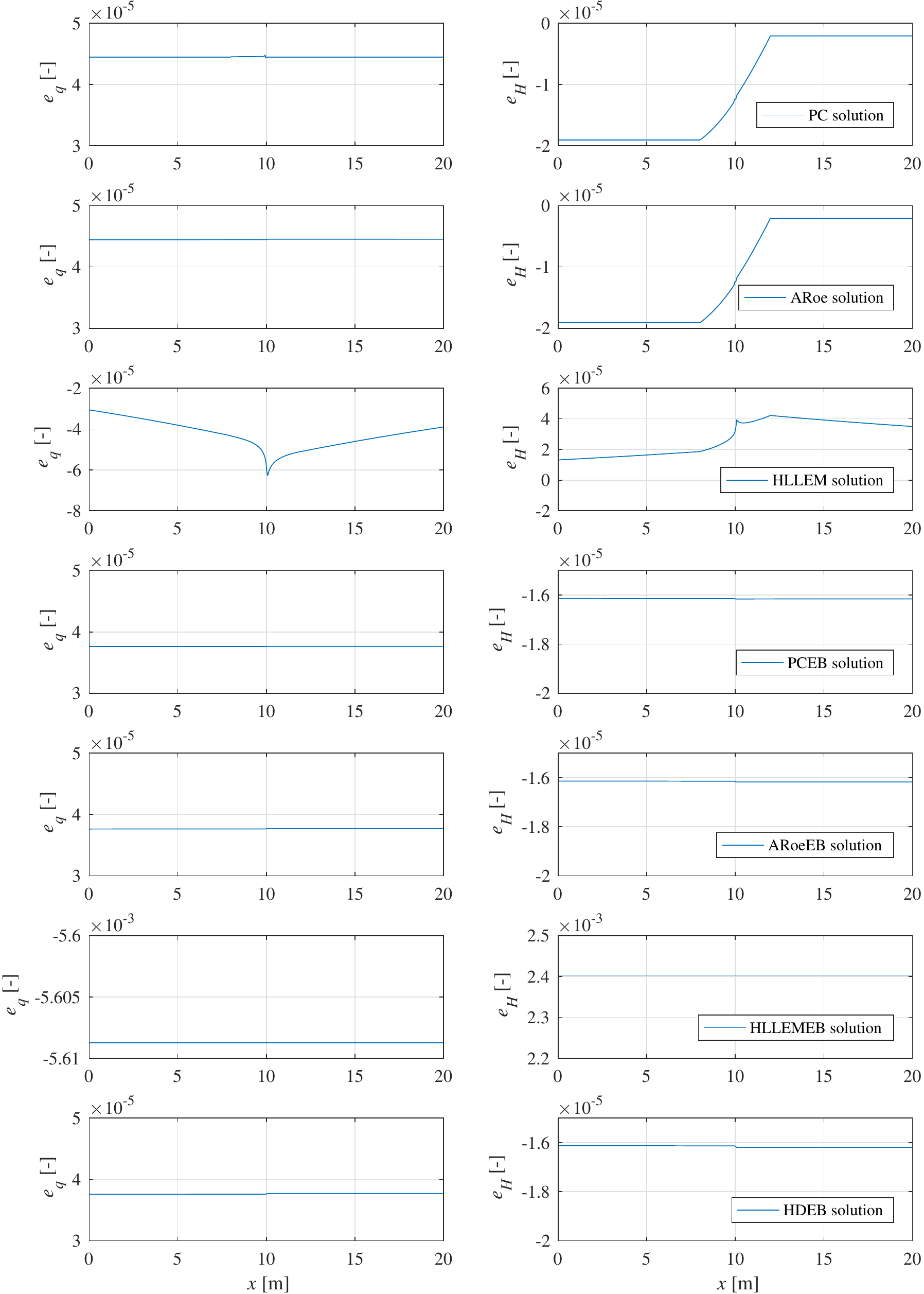}
\end{center}
\caption{Transcritical flow: comparison between errors in the flow discharge and in the total head.}\label{fig:transcritical}
\end{figure}

In Fig.~\ref{fig:transcritical}, the errors $e_q$ and $e_H$ are
reproduced for all the considered models. As shown, all the models are
not able to manage the transition on the critical state. This is
mainly due to the difficulties in developing a consistent numerical
approach near the critical point. In this condition, for a given value
of the total head, it becomes difficult to construct an algorithm to discern the correct physical flow state.

\subsection{Transcritical flow with hydraulic jump over a bump}\label{sec:shockbump}
This test case is devoted to analyzing the influence of the presence
of a hydraulic jump in achieving the energy consistency property of the models.

\begin{figure} 
\begin{center}
\includegraphics[width=\textwidth]{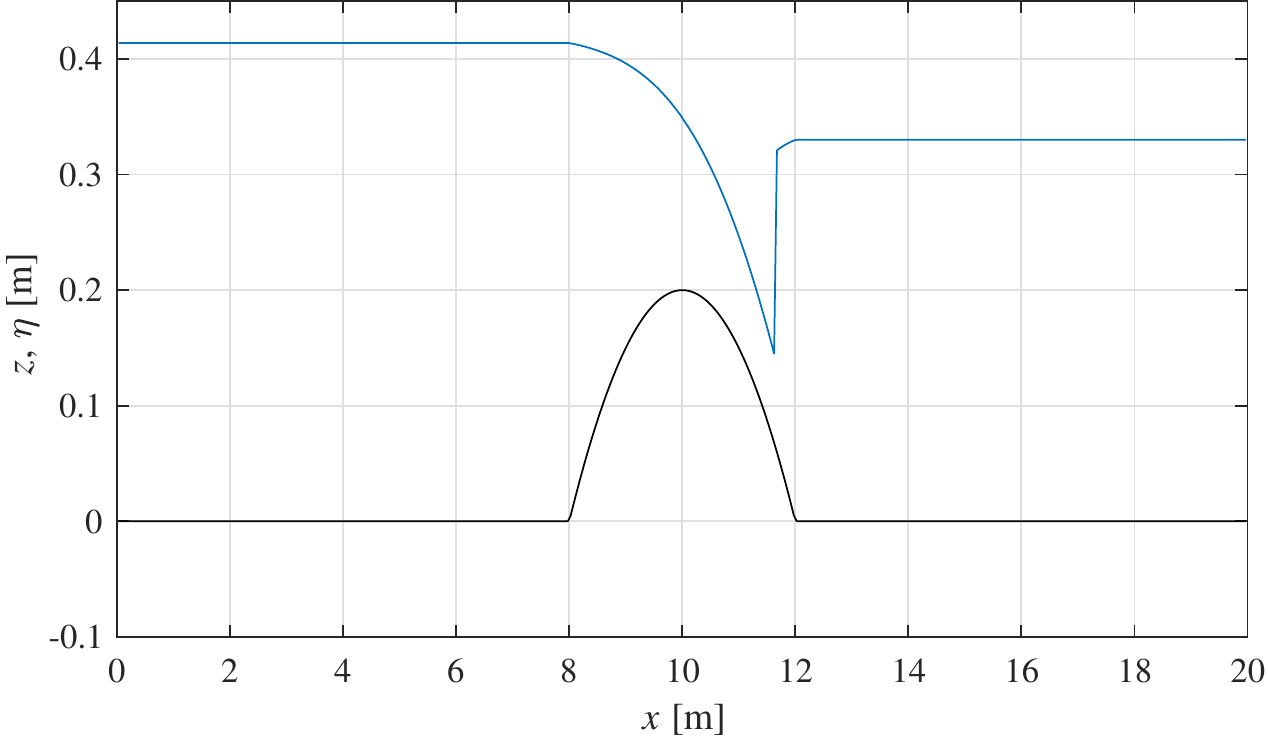}
\end{center}
\caption{Bottom profile and free-surface water level for the flow over
  a bump with a hydraulic jump.}\label{fig:shock_sketch}
\end{figure}

We consider the same channel, bottom profile and space discretization
of the test case presented in section \ref{sec:subsuper}. The upstream specific discharge is 0.18 m$^2$/s, and the downstream water-surface elevation is set to 0.33 m. The initial water level is 0.33 m, and the initial discharge is 0.18 m$^2$/s everywhere. After the initial transient, the asymptotic steady state is reached after 100 s of simulation. The analytical solution is reported in Fig.~\ref{fig:shock_sketch} in terms of free surface elevation.

\begin{figure} 
\begin{center}
\includegraphics[width=\textwidth]{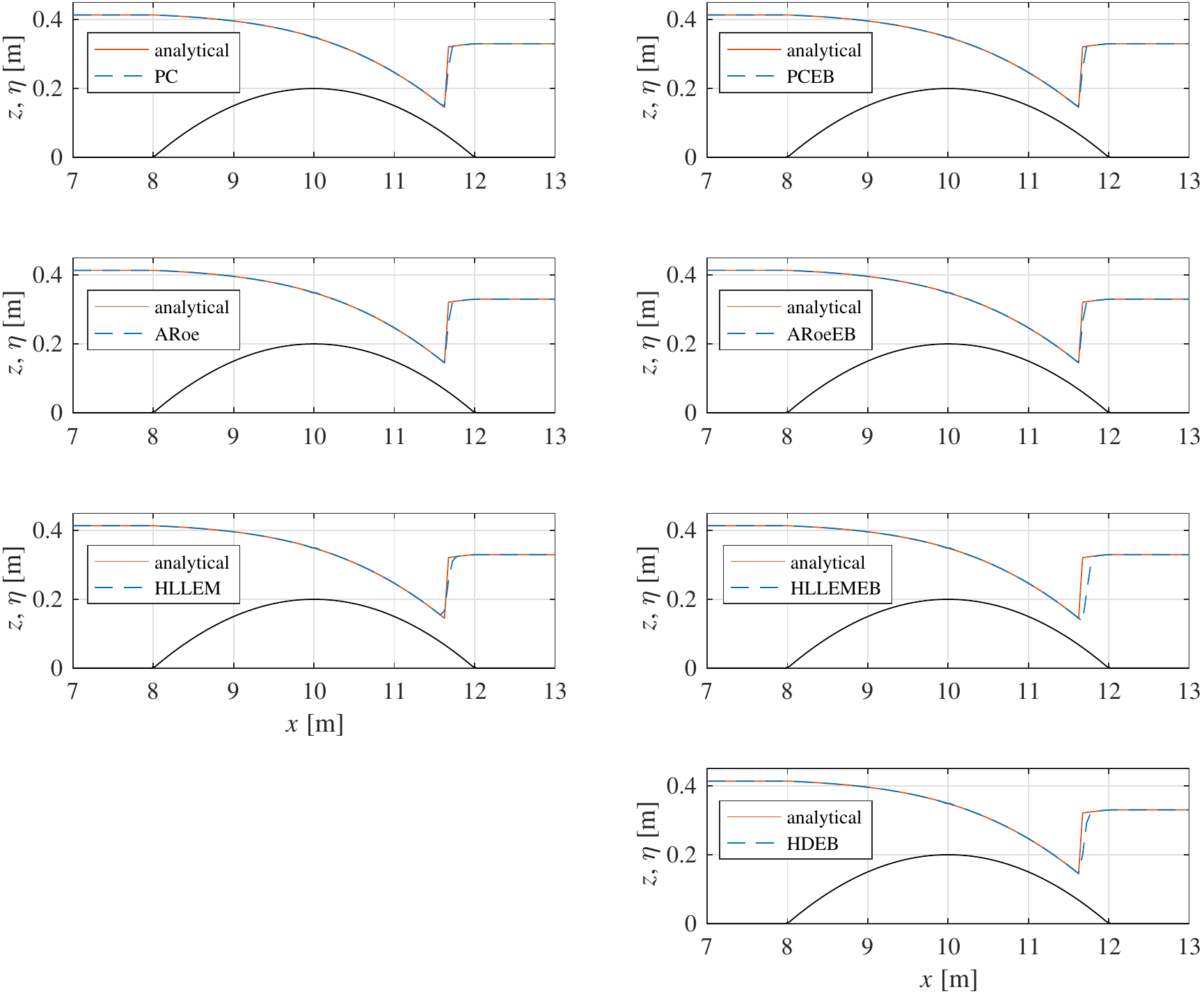}
\end{center}
\caption{Transcritical flow over a bump with a hydraulic jump: comparison between analytical and numerically computed free-surface water level. Only the part of the domain between $x = 7$ m and $x = 13$ m is represented.}\label{fig:shockprof}
\end{figure}

\begin{figure} 
\begin{center}
\includegraphics[width=0.92\textwidth]{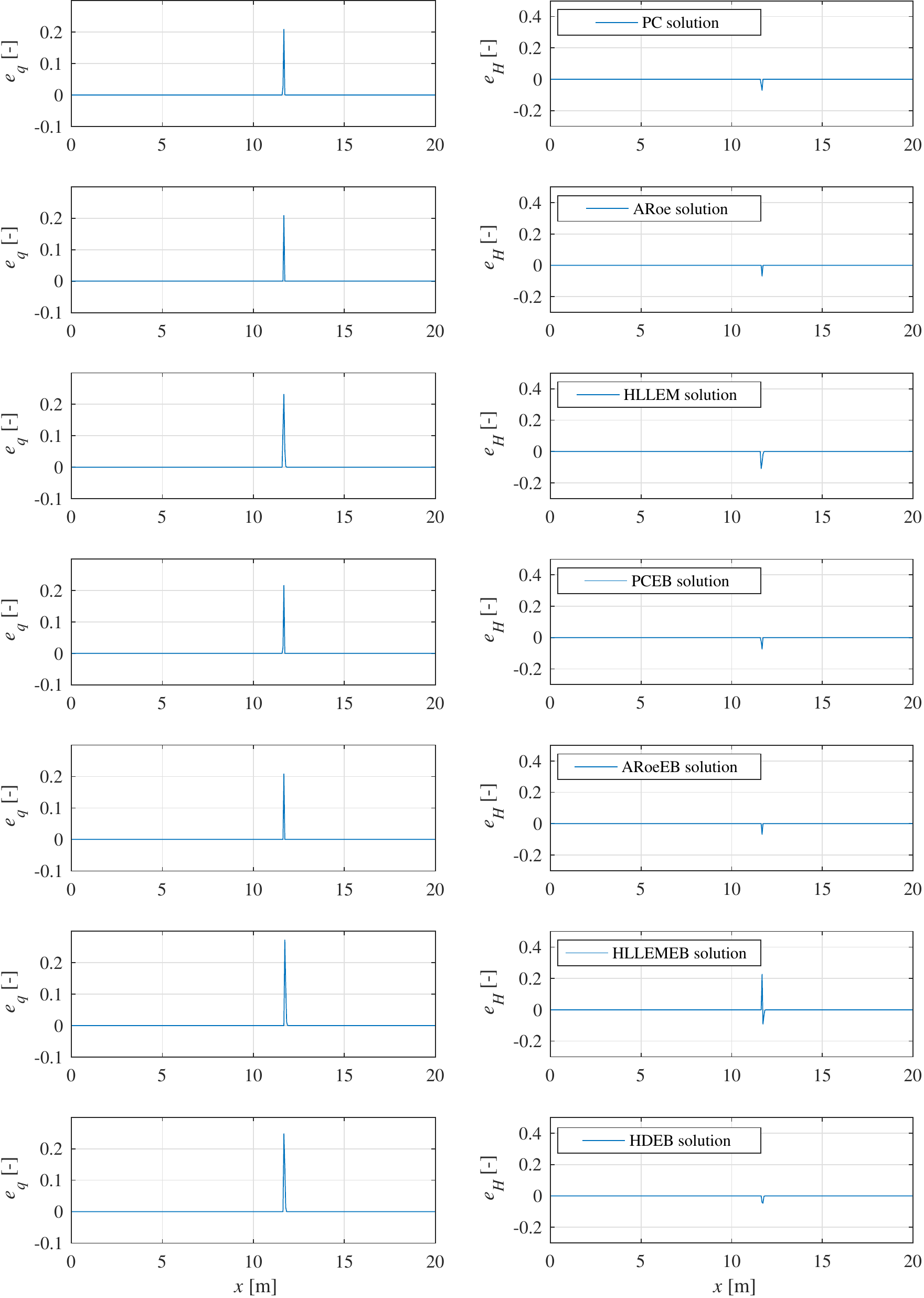}
\end{center}
\caption{Transcritical flow over a bump with a hydraulic jump: comparison between errors in the flow discharge and in the total head.}\label{fig:shock}
\end{figure}

Fig.~\ref{fig:shockprof} describes the solution in terms of water
level for all the considered models, and Fig.~\ref{fig:shock} shows
the errors $e_q$ and $e_H$. As shown, all the models, both the WB and
the EB schemes, are not able to exactly manage the solution
discontinuity constituted by the hydraulic jump. Moreover, all the
models lead to very similar results. The only exception is the HLLEMEB
scheme, which is less accurate in the reproduction of the jump
position. This poor behavior may be due to the well-known problem of
the path-conservative schemes in the reproduction of strong shocks
\cite{Abgrall2010,Castro2008}. In this case, the selection of a
non-linear path leads to an error in the prediction of the jump
position. Notwithstanding, note that in the case of moderate amplitude
shocks, the HLLEMEB model also works properly (see the test case \ref{sec:stoker_step}). The overall performance of all the models is satisfactory, and we can affirm that the technique used to achieve the energy consistency for the smooth solution does not interfere with the behavior of the schemes near the discontinuity.

\subsection{Non-resonant dam-break flow over a step}\label{sec:stoker_step}
\begin{figure} 
\begin{center}
\includegraphics[width=\textwidth]{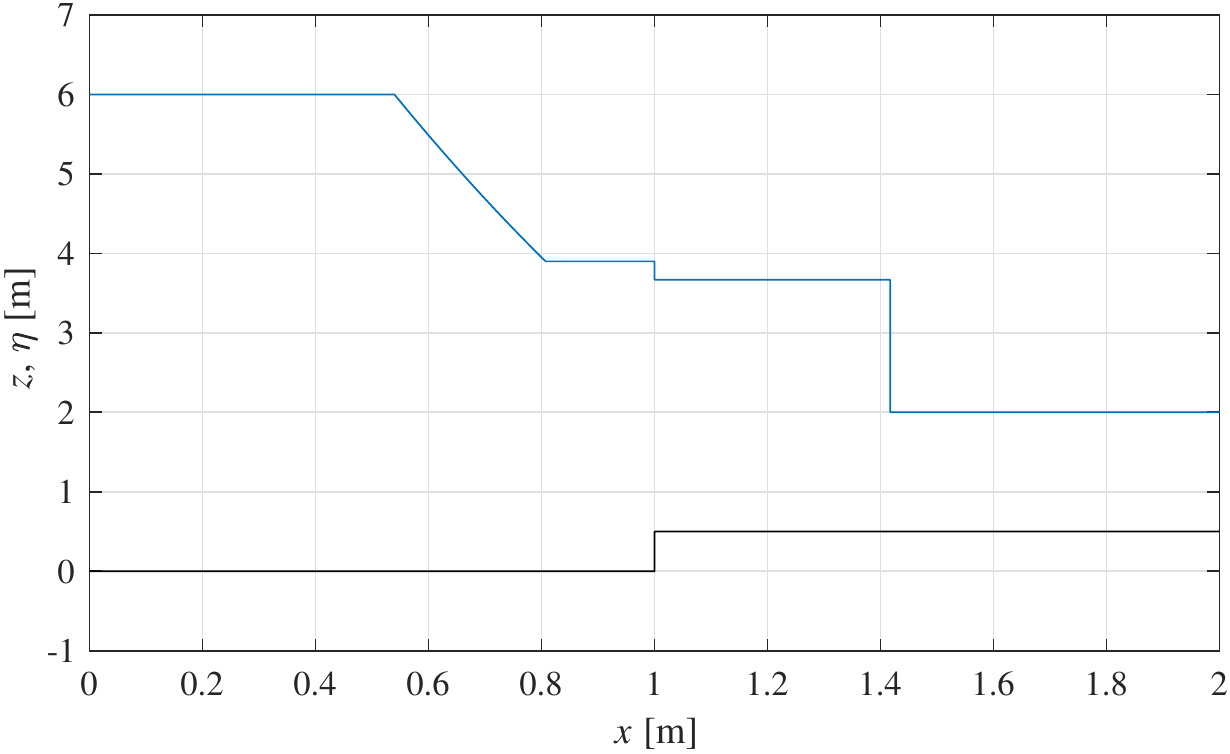}
\end{center}
\caption{Bottom profile and free-surface water level for the non-resonant dam-break flow over a bottom step.}\label{fig:stoker_step_sketch}
\end{figure}

This Riemann problem is proposed in \cite{Caleffi2015a} to verify the
behavior of the numerical models in the reproduction of unsteady
flows. In particular, the ability of the models to reproduce both the
steady discontinuity over the step and the moving hydraulic jump is
tested. The channel is 2 m long with a 0.5 m height forward step at $x
= 1$ m. The initial water level is 6 m for $x < 1$ m and 2 m for $x >
1$ m. The specific discharge is zero in all of the domain. The solution is constituted by a rarefaction, a steady contact wave and a shock. Fig.~\ref{fig:stoker_step_sketch} shows the reference solution of the problem in terms of free-surface elevation \cite{LeFloch2007,LeFloch2011}.

\begin{figure} 
\begin{center}
\includegraphics[width=\textwidth]{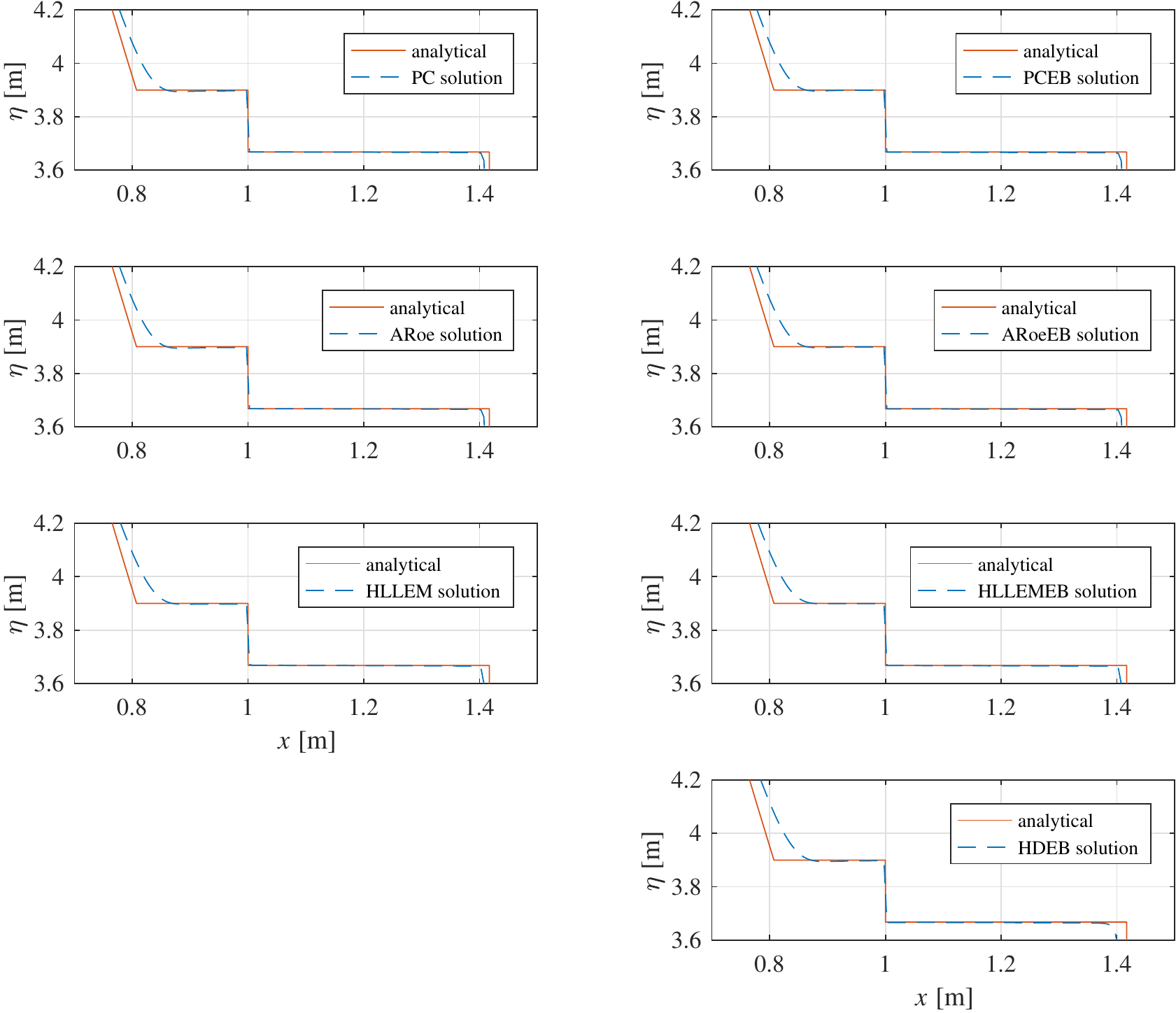}
\end{center}
\caption{Non-resonant dam-break flow over a bottom step: comparison between analytical and numerically computed free-surface water level. Only the part of the domain between $x = 0.7$ m and $x = 1.5$ m is represented.}\label{fig:stoker_step}
\end{figure}

Fig.~\ref{fig:stoker_step} shows the comparison between analytical and
numerical solutions in terms of free-surface elevation. The analysis
of the obtained results allows the conclusion obtained in
\cite{Caleffi2015a} to be confirmed. Both the WB and EB models work well for this test case. 
Moreover, it is well known that the path-conservative models may poorly reproduce the position and the amplitude of a shock \cite{Abgrall2010,Castro2008}, but the presented results make us confident regarding the shock-capturing properties of all the models. This achievement is particularly relevant for the HLLEMEB and PCEB models, where the choice of the non-linear path can raise some perplexities.

\subsection{Resonant Riemann problem over a step}\label{sec:resonant}
The possible discontinuity of the bottom makes the SWE a non-strictly
hyperbolic system of equations. A first consequence of the possible
loss of hyperbolicity is the resonance phenomenon \cite{LeFloch2011}
for certain values of the initial conditions of the Riemann
problem. In these conditions, the elementary waves can interact,
giving rise to a flow that is challenging to be numerically
solved. The  resonant Riemann problem proposed in \cite{LeFloch2011}
is selected here to validate the behavior of the different models in
the case of resonance.

\begin{figure} 
\begin{center}
\includegraphics[width=\textwidth]{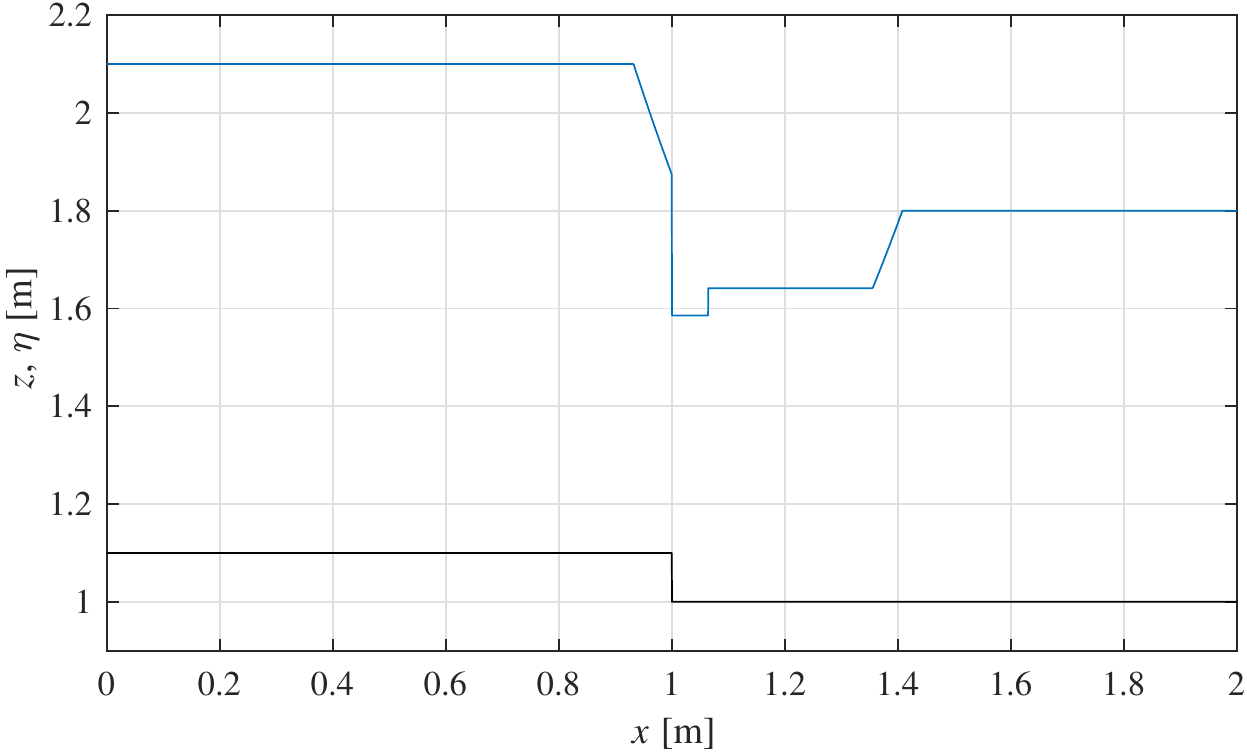}
\end{center}
\caption{Bottom profile and free-surface water level for the resonant Riemann problem.}\label{fig:resonant_sketch}
\end{figure}

The channel is 2 m long. The bottom elevation is 1 m for $x < 1$ m and 1.1 m for $x > 1$ m. The initial water level is 1 m for $x < 1$ m and 0.8 m for $x > 1$ m, and the initial flow velocity is 2 m/s for $x < 1$ m and 4 m/s for $x > 1$ m. The solution is constituted by a rarefaction, a steady contact wave, a shock and a rarefaction. Fig.~\ref{fig:resonant_sketch} shows the reference solution of the problem in terms of free-surface elevation and is computed according to \cite{LeFloch2007,LeFloch2011}.

\begin{figure} 
\begin{center}
\includegraphics[width=\textwidth]{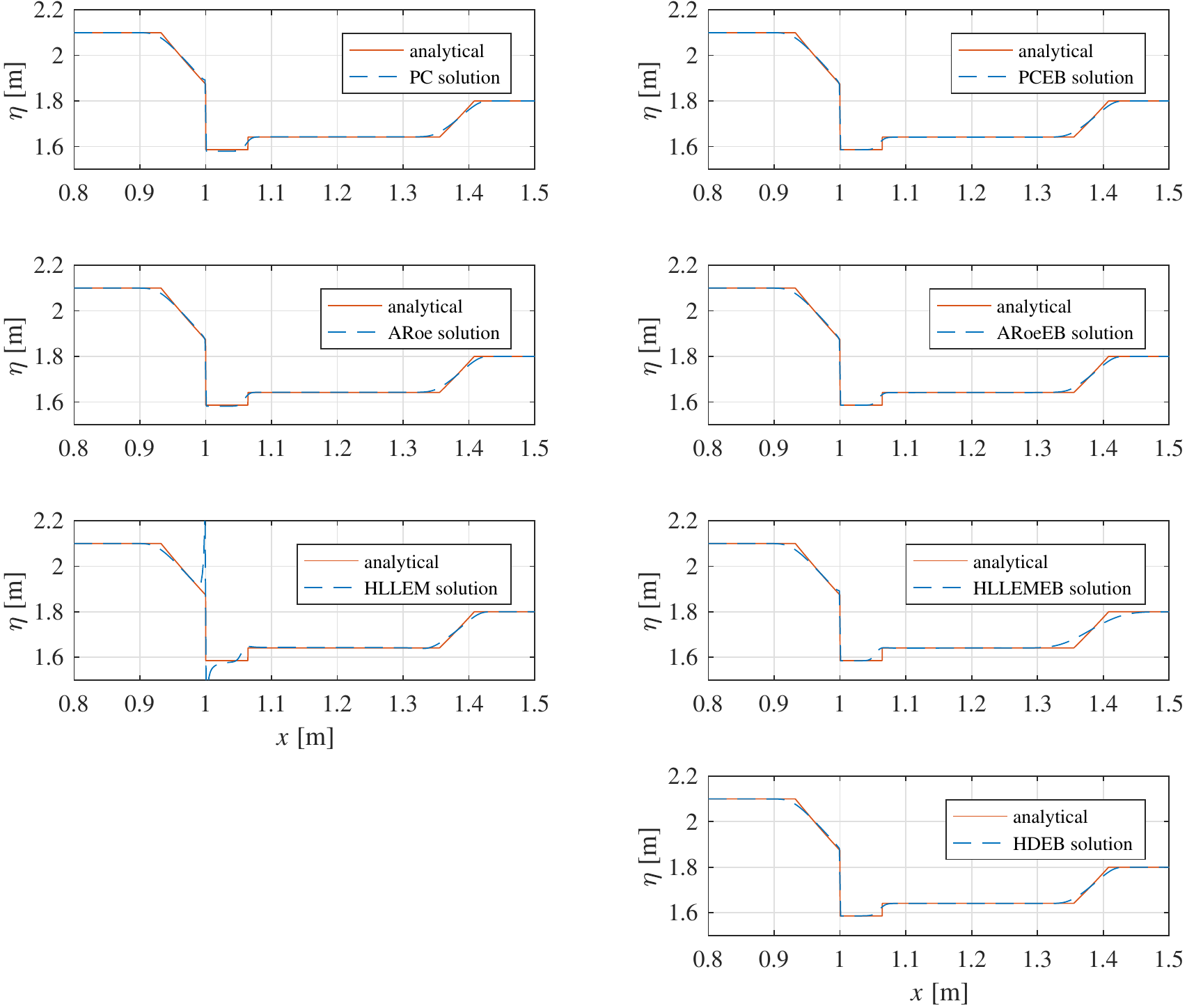}
\end{center}
\caption{Resonant Riemann problem: comparison between analytical and numerically computed free-surface water level. Only the part of the domain between $x = 0.8$ m and $x = 1.5$ m is represented.}\label{fig:resonant_large}
\end{figure}

\begin{figure} 
\begin{center}
\includegraphics[width=\textwidth]{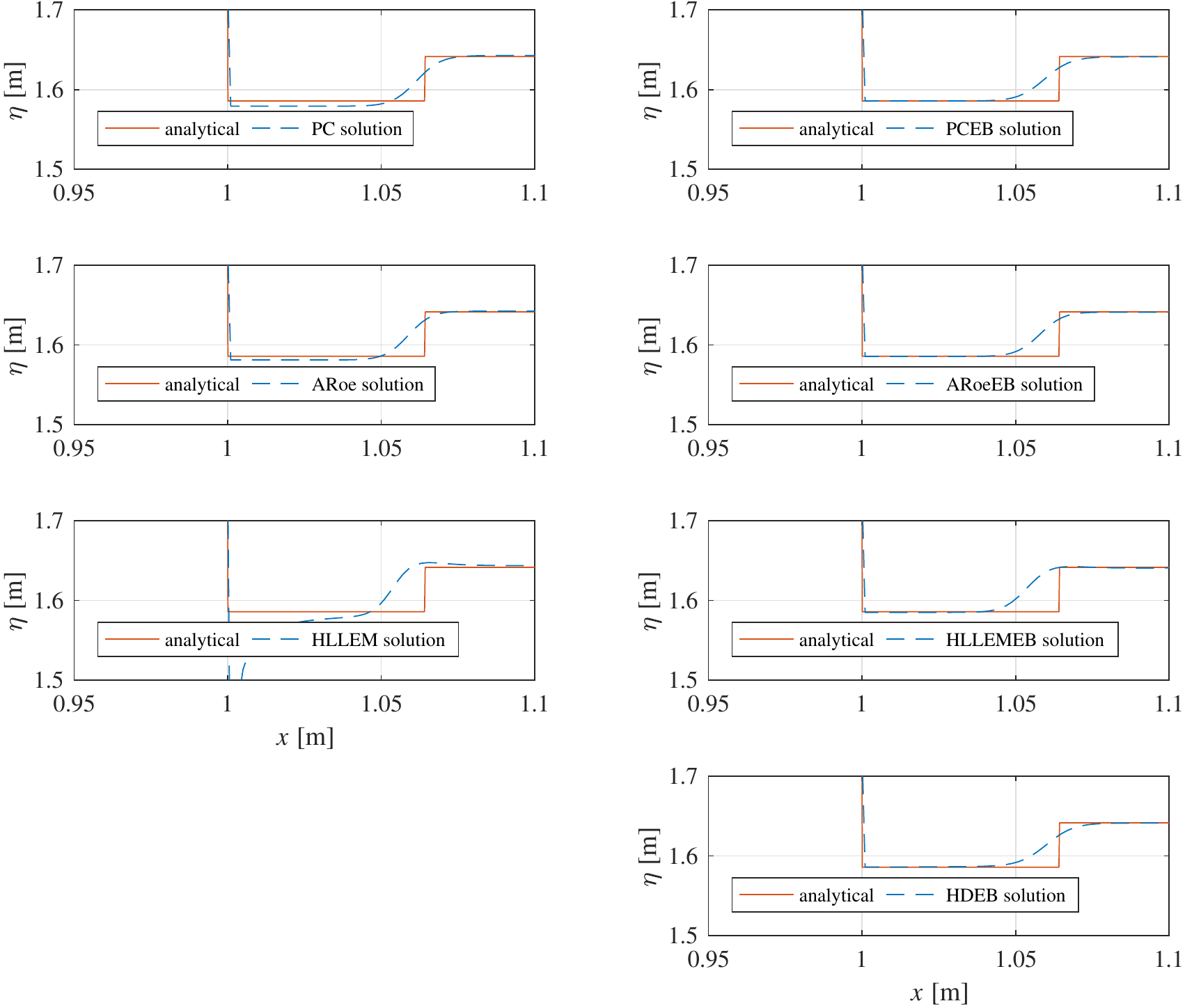}
\end{center}
\caption{Resonant Riemann problem: comparison between analytical and numerically computed free-surface water level. Only the part of the domain between $x = 0.95$ m and $x = 1.1$ m is represented.}\label{fig:resonant_small}
\end{figure}

Figs.~\ref{fig:resonant_large} and \ref{fig:resonant_small} show the
comparison between the analytical and numerical solutions in terms of free-surface elevation. Small portions of the computational domain are represented to highlight the differences between the results.

From Fig.~\ref{fig:resonant_large}, it is possible to observe that the
HLLEM model is not able to reproduce the steady contact discontinuity
at the step. In particular, overshoots and undershoots of the water
elevation near the step are clearly visible. Moreover, the HLLMEB
models required the use of smaller time steps to achieve satisfactory
results. Only in this test case and the HLLEMEB model, the CFL
coefficient is set equal to 0.2 rather than 0.9. Finally, the results related to the PCEB model presented for this test case are obtained using the variant of the model described in \ref{sec:path3s}.

In general, all the models, with the exception of the HLLEM model, are
able to reproduce the analytical solution quite well. Notwithstanding,
the analysis of the solution details reported in
Fig.~\ref{fig:resonant_small} shows that the EB models obtain better
approximations than the WB models. In particular, the amplitude of the free-surface jump at the bottom step is correctly estimated by the EB models and is overestimated by the WB models. This behavior is due to the incorrect energy dissipation introduced at the step by the WB models.
%
\section{Conclusions}
In this work, we present for the first time a new energy-balanced scheme based on a modified version of the HLLEM approximate Riemann solver. Then, we perform a comparison between well-balanced models and energy-balanced models. From this comparison, we are able to highlight the strengths and weaknesses of the two approaches and to show the different behaviors of the different energy-balanced models.

From this study, we can conclude that different classes of well-balanced models can be improved in the reproduction of the asymptotic steady state. In particular, we have considered the path-consistent, the Roe, the HLLEM and the hydrostatic reconstruction schemes. In general, we have found that the accuracy increase in the steady state reproduction is counterbalanced by a reduced robustness and numerical efficiency of the models. We have suggested some solutions to reduce these drawbacks at the cost of increased algorithm complexity.


\section*{Acknowledgments}
This work is funded by: University of Ferrara within the Founding Program FIR 2016, project tile ``Energy-preserving numerical models for the Shallow Water Equations''.

%
\appendix
\section{The explicit expressions for the fluctuations in the PCEB model}\label{apx:explicit}
The direct substitution of the path \eqref{eq:pathnl} in \eqref{eq:Dpm} leads to a time-consuming scheme. In particular, the numerical integration of all the components of the product of the matrices $A$ and $|A|$ with the path derivatives along the path itself is a computationally demanding task. For this reason, in this appendix, we perform some algebraic manipulations to develop a more efficient model.

First, the evaluation of the derivative with respect to $s$ of the three components of the path $\Psi(w^{-},w^{+},s)$ has to be performed. 
The $q$ an $z$ components of the path are linear, and therefore, the
corresponding derivatives are straightforward. The derivative of the
$h$ component is more involved because the depth is not given as a
closed expression of $s$ but rather in an implicit form involving $H$. To address this issue, we consider the total head as a function of $s$:
\begin{equation}
H(s) = z(s) + h(s) + \frac{\left[q(s)\right]^2}{2\,g\,\left[h(s)\right]^2};
\end{equation}
and we perform the following algebraic manipulation:
\begin{multline}
\frac{\partial H}{\partial s} = \frac{\partial z}{\partial s} + \frac{\partial h}{\partial s}
+ \frac{q}{g\,h^2}\frac{\partial q}{\partial s} - 
\frac{q^2}{g\,h^3}\frac{\partial h}{\partial s}; \Rightarrow \\
\left(1 - \frac{q^2}{g\,h^3}\right)\frac{\partial h}{\partial s} = \frac{\partial H}{\partial s} - \frac{\partial z}{\partial s} - \frac{q}{g\,h^2}\frac{\partial q}{\partial s}; \Rightarrow \\
\frac{\partial h}{\partial s} = \left(1 - \frac{q^2}{g\,h^3}\right)^{-1}\left[\frac{\partial H}{\partial s} - \frac{\partial z}{\partial s} - \frac{q}{g\,h^2}\frac{\partial q}{\partial s}\right];
\end{multline}
which leads to the expression of the derivative of $h$:
\begin{equation}\label{eq:dhdsHcostAll}
\frac{\partial h}{\partial s} = 
\left(\frac{c^2}{c^2 - v^2}\right)\left[\frac{\partial H}{\partial s} - \frac{\partial z}{\partial s} - \frac{v}{c^2}\frac{\partial q}{\partial s}\right].
\end{equation}

Taking into account the path \eqref{eq:pathnl}, the linear variation of the total head \eqref{eq:pathHlin} and the relationship \eqref{eq:dhdsHcostAll}, the path derivatives become:
\begin{equation}
\frac{\partial \Psi}{\partial s}= 
\begin{bmatrix}
\partial \bar{h}/\partial s \\
\partial \bar{q}/\partial s \\
\partial \bar{z}/\partial s \\
\end{bmatrix}=
\begin{bmatrix}
\left(\frac{c^2}{c^2 - v^2}\right)\left(\jump{H} - \jump{z} - \frac{v}{c^2}\jump{q}\right) \\
\jump{q}\\
\jump{z}\\
\end{bmatrix}.
\end{equation}

We are now able to write the explicit form of the matrix products that appear in Eq.~\eqref{eq:Dpm}, $A\frac{\partial \Psi}{\partial s}$ and $\left|A\right|\frac{\partial \Psi}{\partial s}$, where $A$ is given in Eq.~\eqref{eq:PCSWE} and $|A| = R\,|\Lambda|\, R^{-1}$.

In the former case, we have:
\begin{equation}
A\frac{\partial \Psi}{\partial s}=
\begin{bmatrix}0&1&0\\c^2-v^2&2v&c^2\\0&0&0\end{bmatrix}
\begin{bmatrix}
\left(\frac{c^2}{c^2 - v^2}\right)\left(\jump{H} - \jump{z} - \frac{v}{c^2}\jump{q}\right) \\
\jump{q}\\
\jump{z}\\
\end{bmatrix};
\end{equation}
which, after a straightforward manipulation, becomes:
\begin{equation}\label{eq:Adpds}
A\frac{\partial \Psi}{\partial s}=
\begin{bmatrix}
\jump{q}\\
v\jump{q} + c^2\jump{H}\\
0
\end{bmatrix}.
\end{equation}

To obtain the explicit expression for $\left|A\right|\frac{\partial \Psi}{\partial s}$, we first introduce an explicit formulation for the matrix $|A|$. With this aim, we write the matrices of the right eigenvectors and its inverse in the form:
\begin{equation}\label{eq:Rexp}
R=\begin{bmatrix}1&\frac{c^2}{v^2 - c^2}&1\\v-c&0&v+c\\0&1&0\end{bmatrix};
\qquad
R^{-1}=\begin{bmatrix} \frac{c+v}{2 c} & -\frac{1}{2 c} & \frac{c}{2 c-2 v} \\
 0 & 0 & 1 \\
 \frac{c-v}{2 c} & \frac{1}{2 c} & \frac{c}{2 (c+v)} \\
\end{bmatrix};
\end{equation}
where the $R_2$ eigenvector of Eq.~\eqref{eq:eigenvectors} is multiplied by $\frac{c^2}{v^2 - c^2}$ for convenience, and the matrix $|\Lambda|$ is:
\begin{equation}\label{eq:absL}
\left|\Lambda\right|=\begin{bmatrix}\left|v-c\right|&0&0\\0&0&0\\0&0&\left|v+c\right|\end{bmatrix}.
\end{equation}

Using Eqs.~\eqref{eq:Rexp} and \eqref{eq:absL}, it is easy to write $|A(w)|= R\,|\Lambda|\, R^{-1}$ in the form:
%
%
\begin{equation}\label{eq:absJexp} \scriptstyle
|A|=\frac{1}{2 c}
\begin{bmatrix}
 \scriptstyle (c+v) \left| v-c\right| + (c-v) \left| c+v\right| &
 \scriptstyle \left| c+v\right| -\left| v-c\right| & 
 \frac{c^2 \left| v-c\right|}{c- v}+\frac{c^2 \left| c+v\right| }{c+v} \\
 \scriptstyle (v^2-c^2) \left(\left| v-c\right| - \left| c+v\right|\right) &
 \scriptstyle (c+v) \left| c+v\right| -(v-c) \left| v-c\right| &
 \frac{c^2 (v-c) \left| v-c\right| }{c-v}+{\scriptstyle c^2 \left| c+v\right| }  \\
 \scriptstyle 0 &\scriptstyle 0 &\scriptstyle 0 \\
\end{bmatrix}.
\end{equation}

\emph{Assuming that $v$ is positive} (i.e., $v+c>0$), we have the following two possibilities.
For a \emph{subcritical flow} (i.e., $v-c<0$), Eq.~\eqref{eq:absJexp} becomes:
%
%
\begin{equation}\label{eq:absA_subcritical}
|A| = \frac{1}{c}
\begin{bmatrix}
 {c^2-v^2} & {v} & c^2 \\
 {c^2 v-v^3} & {v^2+c^2} & c^2 v \\
 0 & 0 & 0 \\
\end{bmatrix};
\end{equation}
and for a \emph{supercritical flow} (i.e., $v-c>0$):
\begin{equation}\label{eq:absA_supercriticalplus}
|A| = 
\begin{bmatrix}
 0 & 1 & 0 \\
c^2-v^2 & 2 v & c^2 \\
 0 & 0 & 0 \\
\end{bmatrix};
\end{equation}

\emph{Assuming that $v$ is negative} (i.e., $v-c<0$), for a \emph{subcritical flow} (i.e., $v+c>0$):
\begin{equation}
|A| = \frac{1}{c}
\begin{bmatrix}
 {c^2-v^2} & {v} & c^2 \\
 {c^2 v-v^3} & {v^2+c^2} & c^2 v \\
 0 & 0 & 0 \\
\end{bmatrix};
\end{equation}
equal to Eq.~\eqref{eq:absA_subcritical}, and for a \emph{supercritical flow} (i.e., $v+c<0$):
\begin{equation}\label{eq:absA_supercriticalminus}
|A| = -
\begin{bmatrix}
 0 & 1 & 0 \\
c^2-v^2 & 2 v & c^2 \\
 0 & 0 & 0 \\
\end{bmatrix}.
\end{equation}
In other words, in the case of a subcritical flow, $|A|$ is given by
Eq.~\eqref{eq:absA_subcritical}; in case of a supercritical flow with
$v>0$, $|A|=A$; and in the case of a supercritical flow with $v<0$, $|A|=-A$.

Therefore, taking into account Eq.~\eqref{eq:absA_subcritical}, in the
case of a subcritical flow, we have:
\begin{equation}
|A| \frac{\partial \Psi}{\partial s}= 
\frac{1}{c}
\begin{bmatrix}
 {c^2-v^2} & {v} & c^2 \\
 {c^2 v-v^3} & {v^2+c^2} & c^2 v \\
 0 & 0 & 0 \\
\end{bmatrix}
\begin{bmatrix}
\left(\frac{c^2}{c^2 - v^2}\right)\left(\jump{H} - \jump{z} - \frac{v}{c^2}\jump{q}\right) \\
\jump{q}\\
\jump{z}\\
\end{bmatrix};
\end{equation}
%
%
and, after some manipulations:
\begin{equation}\label{eq:absAsub}
|A| \frac{\partial \Psi}{\partial s}= 
\begin{bmatrix}
c\jump{H}\\
c \jump{q} + {c v}\jump{H}\\
 0 \\
\end{bmatrix};
\end{equation}

Whereas in the case of a supercritical flow, we have:
\begin{equation}\label{eq:absAsup}
|A| \frac{\partial \Psi}{\partial s}= 
\begin{bmatrix}
\jump{q}\\
v\jump{q} + c^2\jump{H}\\
0
\end{bmatrix},\quad\text{and}\quad
|A| \frac{\partial \Psi}{\partial s}= 
- \begin{bmatrix}
\jump{q}\\
v\jump{q} + c^2\jump{H}\\
0
\end{bmatrix};
\end{equation}
for $v>0$ and $v<0$, respectively.

Taking into account the Dumbser-Osher-Toro (DOT) Riemann solver \cite{DOT} of Eq.~\eqref{eq:Dpm}:
\begin{equation}
\mathcal{D}^{\pm} = \frac{1}{2}\int_0^1 \left(
A \pm 
\left|A\right|\right)\frac{\partial \Psi}{\partial s}\ \d s
\end{equation}
and Eqs.~\eqref{eq:Adpds}, \eqref{eq:absAsub} and \eqref{eq:absAsup}, for the subcritical case:
\begin{equation}\label{eq:Dsub}
\mathcal{D}^{\pm} = \frac{1}{2}\int_0^1 
\begin{bmatrix}
\jump{q} \pm c \jump{H}\\
(v \pm c) \jump{q} + c(c\pm v)\jump{H}\\
0 \\
\end{bmatrix} \ \d s
\end{equation}
and for the supercritical case ($v>0$):
\begin{equation}\label{eq:Dsupplus}
\mathcal{D}^{\pm} = \left(\frac{1}{2}\pm\frac{1}{2}\right)\int_0^1 
\begin{bmatrix}
\jump{q} \\
v \jump{q} + c^2\jump{q}\\
0 \\
\end{bmatrix} \ \d s
\end{equation}
and for the supercritical case ($v<0$):
\begin{equation}\label{eq:Dsupminus}
\mathcal{D}^{\pm} = \left(\frac{1}{2}\mp\frac{1}{2}\right)\int_0^1 
\begin{bmatrix} 
\jump{q} \\
v \jump{q} + c^2\jump{H}\\
0 \\
\end{bmatrix} \ \d s
\end{equation}

Finally, Eqs.~\eqref{eq:Adpds}-\eqref{eq:Dsupminus} can be also written as:
\begin{equation}\label{eq:flucfinal1}
\mathcal{D}^{\pm} =
\begin{bmatrix}
\frac{1}{2}\jump{q} \pm \frac{1}{2}\jump{H} \int_0^1 c\ \d s \\
\frac{1}{2}\jump{q} \int_0^1 (v \pm c) \ \d s + \frac{1}{2}\jump{H} \int_0^1 c(c\pm v)\ \d s\\
0 \\
\end{bmatrix}
\end{equation}
for a subcritical flow, and:
\begin{equation}
\mathcal{D}^{+} = 
\begin{bmatrix}
\jump{q}\\
\jump{q}\int_0^1 v \ \d s + \jump{H}\int_0^1 c^2\ \d s\\
0 \\
\end{bmatrix}; \qquad 
\mathcal{D}^{-} = 
\begin{bmatrix}
0 \\
0 \\
0 \\
\end{bmatrix};
\end{equation}
for a supercritical flow with ($v>0$), and: 
\begin{equation}\label{eq:flucfinal3}
\mathcal{D}^{+} = 
\begin{bmatrix}
0 \\
0 \\
0 \\
\end{bmatrix};\qquad 
\mathcal{D}^{-} = 
\begin{bmatrix}
\jump{q}\\
\jump{q}\int_0^1 v \ \d s + \jump{H}\int_0^1 c^2\ \d s\\
0 \\
\end{bmatrix}; 
\end{equation}
for a supercritical flow with ($v<0$).

Clearly, Eqs.~\eqref{eq:flucfinal1}-\eqref{eq:flucfinal3} allow a very efficient implementation of the DOT Riemann solver after the numerical evaluation of the integrals of the scalar quantities, $c$, $c^2$, $v$ and $cv$.
\section{The integration path for the transcritical flow in the PCEB model}\label{sec:path3s}
The use of the non-linear path proposed in \S~\ref{sec:pcn} performs
poorly in the case of transcritical flow (i.e., the flow is
subcritical for $w^{-}$ and supercritical for $w^{+}$ or
\emph{vice-versa}). For this reason, we propose a modified path to be
inserted into Eq.~\eqref{eq:Dpm} in the case of transcritical flows. The key idea is to split the path into three parts: one subcritical, one supercritical and one connecting the two.

The first step is the location of a coordinate, $s_0$, along the path that corresponds to the critical state characterized by a Froude number equal to 1 (i.e., $\ten{Fr}(s_0)=1$). To this end, we assume that the function $\ten{Fr}(s)$ is given by:
\begin{equation}
\ten{Fr}(s) = \frac{q(s)}{h(s)\sqrt{g\,h(s)}};
\end{equation}
with:
\begin{equation}
h(s) = h^- + s (h^+ - h^-); \quad \text{and} \quad
q(s) =  q^- + s (q^+ - q^-);
\end{equation}
and afterward, we  numerically solve the relationship $\ten{Fr}(s_0)=1$.

With $s_0$ available, we can calculate:
\begin{align}
z_0 &= z^- + s_0 (z^+ - z^-); \\
q_0 &= q^- + s_0 (q^+ - q^-); \\
H_0 &= H^- + s_0 (H^+ - H^-); \\
E_0 &= H_0 - z_0;
\end{align}
and also $h_0^-$ and $h_0^+$, given by:
\begin{align}
h_0^- &= E_0^{-1};& \text{with $h_0^-$ subcritical if $\ten{Fr}^- < 1$ else $h_0^-$ supercritical};\\
h_0^+ &= E_0^{-1};& \text{with $h_0^+$ subcritical if $\ten{Fr}^+ < 1$ else $h_0^+$ supercritical}.
\end{align}

The computation of $\bar{E_0}^{-1}$, i.e., finding the values of $h_0^{\pm}$ that satisfy $h_0^{\pm}+{q_0^2}/\{2\,g\,[h_0^{\pm}]^2\} = E_0$, is analytically computed using the solution given in \cite{Valiani2008}.

The following 4 points in the phase space are considered:
\begin{equation}
w^- = \begin{bmatrix} h^- \\ q^- \\ z^-\end{bmatrix} \quad w^-_0 = \begin{bmatrix} h^-_0 \\ q_0 \\ z_0\end{bmatrix} \quad w^+_0 = \begin{bmatrix} h^+_0 \\ q_0 \\ z_0\end{bmatrix} \quad w^+ = \begin{bmatrix} h^+ \\ q^+ \\ z^+\end{bmatrix};
\end{equation}
which define the path in three parts as $w^- \rightarrow w^-_0 \rightarrow w^+_0 \rightarrow w^+$. The intermediate part is assumed to be linear, whereas the other parts are constructed following the idea proposed in \S~\ref{sec:pcn}. Moreover, indicating with $\Delta w_1$, $\Delta w_2$, $\Delta w_3$ the Euclidean distances between the four points, $w^-$, $w^-_0$, $w^+_0$ and $w^+$, the following curvilinear abscissa can be found: 
\begin{equation}
s_1 = \frac{\Delta w_1}{\sum_{p=1}^3 \Delta w_p}; \qquad s_2 = \frac{\Delta w_1 + \Delta w_2}{\sum_{p=1}^3 \Delta w_p}.
\end{equation}

The path is given by:
\begin{equation}\label{eq:pathAll}
\Psi(s) = \begin{bmatrix}
\bar{h}(s)   \\ 
\bar{q}(s)   \\
\bar{z}(s)   \end{bmatrix};
\end{equation}
with:
\begin{equation}\label{eq:pathz}
\bar{z}(s)=\left\{\begin{array}{lll}
z^- + s \frac{z_0 - z^-}{s_1} & \text{if} & 0 \leq s \leq s_1;\\[1ex]
z_0 & \text{if} & s_1 < s < s_2;\\[1ex]
\frac{z_0 - s_2 \, z^+}{1-s_2} + s \frac{z^+ - z_0}{1-s_2} & \text{if} & s_2 \leq s \leq 1;
\end{array}\right.
\end{equation}
and:
\begin{equation}\label{eq:pathq}
\bar{q}(s)=\left\{\begin{array}{lll}
q^- + s \frac{q_0 - q^-}{s_1} & \text{if} & 0 \leq s \leq s_1;\\[1ex]
q_0 & \text{if} & s_1 < s < s_2;\\[1ex]
\frac{q_0 - s_2 \, q^+}{1-s_2} + s \frac{q^+ - q_0}{1-s_2} & \text{if} & s_2 \leq s \leq 1;
\end{array}\right.
\end{equation}
which corresponds to linear bottom elevation and specific discharge.

According to the procedure proposed in \S~\ref{sec:pcn}, the computation of $\bar{h}$ is slightly more complex. First, the total head along the path is defined as:
\begin{equation}\label{eq:pathH}
\bar{H}(s)=\left\{\begin{array}{lll}
H^- + s \frac{H_0 - H^-}{s_1} & \text{if} & 0 \leq s \leq s_1;\\[1ex]
H_0 & \text{if} & s_1 < s < s_2;\\[1ex]
\frac{H_0 - s_2 \, H^+}{1-s_2} + s \frac{H^+ - H_0}{1-s_2} & \text{if} & s_2 \leq s \leq 1;
\end{array}\right.
\end{equation}
where we compute $\bar{E} = \bar{H} -\bar{z}$, and finally, $\bar{h}$ is given by:
\begin{equation}\label{eq:pathh}
\bar{h}(s)=\left\{\begin{array}{lll}
\bar{E}^{-1}; \text{$\bar{h}$ subcritical if $\ten{Fr}^- < 1$ else $\bar{h}$ supercritical}; & \text{if} & 0 \leq s \leq s_1;\\[2.5ex]
h_0^- + \frac{h_0^+ - h_0^-}{s_2 - s_1} \left(s-s_1\right) & \text{if} & s_1 < s < s_2;\\[3.0ex]
\bar{E}^{-1}; \text{$\bar{h}$ subcritical if $\ten{Fr}^+ < 1$ else $\bar{h}$ supercritical}; & \text{if} & s_2 \leq s \leq 1;
\end{array}\right.
\end{equation}

The derivative of the path with respect to $s$:
\begin{equation}\label{eq:pathderAll}
\frac{\partial \Psi}{\partial s}(s) = \begin{bmatrix}
\frac{\partial \bar{h}}{\partial s}(s)   \\[2.5ex] 
\frac{\partial \bar{q}}{\partial s}(s)   \\[2.5ex]
\frac{\partial \bar{z}}{\partial s}(s)   \end{bmatrix};
\end{equation}
follows from the path definition. Taking into account Eqs.~\eqref{eq:pathz}-\eqref{eq:pathq}, it is easy to write:
\begin{equation}\label{eq:pathdz}
\frac{\partial \bar{z}}{\partial s}(s)=\left\{\begin{array}{lll}
\frac{z_0 - z^-}{s_1} & \text{if} & 0 \leq s \leq s_1;\\[1ex]
0 & \text{if} & s_1 < s < s_2;\\[1ex]
\frac{z^+ - z_0}{1-s_2} & \text{if} & s_2 \leq s \leq 1;
\end{array}\right.
\end{equation}
\begin{equation}\label{eq:pathdq}
\frac{\partial \bar{q}}{\partial s}(s)=\left\{\begin{array}{lll}
\frac{q_0 - q^-}{s_1} & \text{if} & 0 \leq s \leq s_1;\\[2.5ex]
0 & \text{if} & s_1 < s < s_2;\\[1ex]
\frac{q^{+} - q_0}{1-s_2} & \text{if} & s_2 \leq s \leq 1;
\end{array}\right.
\end{equation}
and, taking into account the relationship \eqref{eq:dhdsHcostAll}:
\begin{equation}\label{eq:pathdh}
\frac{\partial \bar{h}}{\partial s}(s)=\left\{\begin{array}{lll}
\left(\frac{c^2}{c^2 - v^2}\right)\left[\frac{\partial \bar{H}}{\partial s} - \frac{\partial \bar{z}}{\partial s} - \frac{\bar{q}}{g\,\bar{h}^2}\frac{\partial \bar{q}}{\partial s}\right] & \text{if} & 0 \leq s \leq s_1;\\[2.5ex]
\frac{h_0^+ - h_0^-}{s_2 - s_1} & \text{if} & s_1 < s < s_2;\\[1ex]
\left(\frac{c^2}{c^2 - v^2}\right)\left[\frac{\partial \bar{H}}{\partial s} - \frac{\partial \bar{z}}{\partial s} - \frac{\bar{q}}{g\,\bar{h}^2}\frac{\partial \bar{q}}{\partial s}\right] & \text{if} & s_2 \leq s \leq 1;
\end{array}\right.
\end{equation}
with:
\begin{equation}\label{eq:pathdH}
\frac{\partial \bar{H}}{\partial s}(s)=\left\{\begin{array}{lll}
\frac{H_0 - H^-}{s_1} & \text{if} & 0 \leq s \leq s_1;\\[1ex]
0 & \text{if} & s_1 < s < s_2;\\[1ex]
\frac{H^+ - H_0}{1-s_2} & \text{if} & s_2 \leq s \leq 1;
\end{array}\right.
\end{equation}

Finally, Eq.~\eqref{eq:Dpm} is written as:
\begin{multline} \label{eq:Dpm3p}
\mathcal{D}^{\pm} = \frac{1}{2}\int_0^{s_1} \left[
A(\Psi(w^{-},w^{+},s)) \pm 
\left|A(\Psi(w^{-},w^{+},s)) \right|\right]\frac{\partial \Psi}{\partial s}\ \d s + \\
+\frac{1}{2}\int_{s_1}^{s_2} \left[
A(\Psi(w^{-},w^{+},s)) \pm 
\left|A(\Psi(w^{-},w^{+},s)) \right|\right]\frac{\partial \Psi}{\partial s}\ \d s + \\
+\frac{1}{2}\int_{s_2}^{1} \left[
A(\Psi(w^{-},w^{+},s)) \pm
\left|A(\Psi(w^{-},w^{+},s)) \right|\right]\frac{\partial \Psi}{\partial s}\ \d s;
\end{multline} 
with the path, $\Psi$, and the path derivative, ${\partial \Psi}/{\partial s}$, given by \eqref{eq:pathAll} and \eqref{eq:pathderAll}, respectively.



\clearpage

\end{document}